%% file: main.tex
\documentclass[10pt,journal]{IEEEtran}
\usepackage{amsmath,amssymb}
\usepackage{mathtools}  % \bigtimes
\usepackage{graphicx}
\usepackage{xspace}
\usepackage{multirow}
\usepackage{booktabs}
\usepackage[flushleft,para]{threeparttable}
\usepackage{siunitx}
\usepackage[labelformat=simple]{subcaption}
\usepackage[ruled,linesnumbered]{algorithm2e} % algo pseudo code package
	
\usepackage{dcolumn}
	\newcolumntype{P}[1]{D{.}{.}{#1}} % column with given number of decimal places
\usepackage{macros}
\usepackage{hyperref}
\usepackage{tikz}
\usepackage{pgfplots}
	\pgfplotsset{compat=1.14}
\usepackage{mathdots}
\usepackage{color}
\usepackage{xifthen}
\usetikzlibrary{fadings}
\usetikzlibrary{patterns}
\usetikzlibrary{shadows.blur}
\usetikzlibrary{shapes}
\newtheorem{proposition}{Proposition}

\begin{document}
\title{Reachability analysis of linear hybrid systems \\ via block decomposition}

% If you use the hyperref package, please uncomment the following line
% to display URLs in blue roman font according to Springer's eBook style:
% \renewcommand\UrlFont{\color{blue}\rmfamily}
\hyphenation{op-tical net-works semi-conduc-tor}

\author{Sergiy Bogomolov,
        Marcelo Forets,
        Goran Frehse,
        Kostiantyn Potomkin, and
        Christian Schilling%
\thanks{This research was supported in part by the Austrian Science Fund (FWF) under grants S11402-N23 (RiSE/SHiNE) and Z211-N23 (Wittgenstein Award), the European Union's Horizon 2020 research and innovation programme under the Marie Sk{\l}odowska-Curie grant agreement \mbox{No.\ 754411}, and the Air Force Office of Scientific Research under award number \mbox{FA2386-17-1-4065}. Any opinions, findings, and conclusions or recommendations expressed in this material are those of the authors and do not necessarily reflect the views of the United States Air Force.
This is a preprint of an article presented in the International Conference on Embedded Software 2020 that will appear as part of the ESWEEK-TCAD special issue.}
\thanks{S. Bogomolov is with Newcastle University, United Kingdom, (e-mail: sergiy.bogomolov@newcastle.ac.uk).}
\thanks{M. Forets is with Universidad de la Rep\'ublica, CURE, Maldonado, Uruguay, (e-mail: mforets@gmail.com).}%
\thanks{G. Frehse is with ENSTA ParisTech - U2IS, Palaiseau Cedex, France, (e-mail: goran.frehse@ensta-paris.fr).}%
\thanks{K. Potomkin is with Newcastle University, United Kingdom, (e-mail: K.Potomkin2@newcastle.ac.uk).}
\thanks{C. Schilling is with IST Austria, Klosterneuburg, Austria, (e-mail: christian.schilling@ist.ac.at).}%
}

\markboth{Reachability analysis of linear hybrid systems via block decomposition}%
% {Bogomolov \MakeLowercase{\textit{et al.}}: Reachability analysis via block decomposition}
{Reachability analysis via block decomposition}

\IEEEtitleabstractindextext{%
\begin{abstract}
\input{abstract}
\end{abstract}

\begin{IEEEkeywords}
Reachability, Hybrid systems, Decomposition.
\end{IEEEkeywords}
}

\maketitle

\IEEEdisplaynontitleabstractindextext

% \fi
\IEEEpeerreviewmaketitle

\input{introduction}

\input{preliminaries}

\input{algorithm_naive}

\input{algorithm_new}

\input{evaluation}

\input{conclusion}

\bibliographystyle{IEEEtran}
\bibliography{bibliography}

\input{proofs}

\end{document}

%% file: abstract.tex
Reachability analysis aims at identifying states reachable by a system within a given time horizon. This task is known to be computationally expensive for linear hybrid systems. Reachability analysis works by iteratively applying continuous and discrete post operators to compute states reachable according to continuous and discrete dynamics, respectively. In this paper, we enhance both of these operators and make sure that most of the involved computations are performed in low-dimensional state space. In particular, we improve the continuous-post operator by performing computations in high-dimensional state space only for time intervals relevant for the subsequent application of the discrete-post operator. Furthermore, the new discrete-post operator performs low-dimensional computations by leveraging the structure of the guard and assignment of a considered transition. We illustrate the potential of our approach on a number of challenging benchmarks.

%% file: introduction.tex
\section{Introduction}

\IEEEPARstart{A}{} hybrid system~\cite{Henzinger96} is a formalism for modeling cyber-physical systems.
Reachability analysis is a rigorous way to reason about the behavior of hybrid systems.

In this paper, we describe a new reachability algorithm for linear hybrid systems, i.e., hybrid systems with dynamics given by linear differential equations and invariants and guards given by linear inequalities.
The key feature of our algorithm is that it performs computations in low-dimensional subspaces, which greatly improves scalability.
To this end, we integrate (and modify) a recent reachability algorithm for (purely continuous) LTI systems~\cite{BogomolovFFVPS18}, which we call \dwelldeco in the following, in a new algorithm for linear hybrid systems.

The \dwelldeco algorithm decomposes the calculation of the reachable states into calculations in subspaces (called \emph{blocks}).
This decomposition has two benefits.
The first benefit is that computations in lower dimensions are generally more efficient and thus the algorithm is highly scalable.
The second benefit is that the analysis for different subspaces is decoupled; hence one can effectively skip the computations for dimensions that are of no interest (e.g., for a safety property).

Conceptually, reachability analysis for hybrid systems is performed in a ``hybrid loop'' that interleaves a continuous-post algorithm and a discrete-post algorithm.
If we consider \dwelldeco as a black box, we can plug it into this hybrid loop, which we refer to as \hybridalgo (cf.\ Section~\ref{sec:reach_naive}).
However, there are two shortcomings of such a naive integration.
First, all operations besides \dwelldeco are still performed in high dimensions, making \hybridalgo still suffer from scalability issues.
Second, \hybridalgo does not utilize the decoupling of \dwelldeco at all.

We demonstrate that, unlike in \hybridalgo, it is possible to perform all computations in low dimensions (cf.\ Section~\ref{sec:reach_new}).
For this purpose, we modify \dwelldeco as well as \hybridalgo.
In common cases, our algorithm does not introduce an additional approximation error.
Furthermore, our algorithm makes proper use of the second benefit of \dwelldeco by computing the reachable states only in specific dimensions whenever possible.

We implemented the algorithm in JuliaReach, a toolbox for reachability analysis~\cite{BogomolovFFPS19,JuliaReach}, and we evaluate the algorithm on several benchmark problems, including a 1024-dimensional hybrid system (cf.\ Section~\ref{sec:evaluation}).
Our algorithm outperforms the naive \hybridalgo and other state-of-the-art algorithms by several orders of magnitude.

\medskip

To summarize, we show how to modify the decomposed reachability algorithm for LTI systems from~\cite{BogomolovFFVPS18} in order to efficiently integrate it into a new, decomposed reachability algorithm for linear hybrid systems.
The key insights are (1)~to exploit the decomposed structure of the algorithm to perform all operations in low dimensions and (2)~to only compute the reachable states in specific dimensions whenever possible.

\subsection*{Related work}

\paragraph*{Reachability analysis for linear hybrid systems}

There are several tools available for the analysis of the class of systems we consider in this paper.
All participating tools in a recent friendly competition~\cite{althoff2019arch} follow the architecture of two post operators, one for continuous time and one for discrete time~\cite{Althoff15,SchuppAMK17,BakD17,BogomolovFFPS19,FrehseLGDCRLRGDM2011,GurungRBBG19}.
We implemented our algorithm in JuliaReach~\cite{BogomolovFFPS19} and will consider the mature tool SpaceEx~\cite{FrehseLGDCRLRGDM2011} in the evaluation in Section~\ref{sec:evaluation}.

\paragraph*{Decomposition}

Hybrid systems given as a network of components can be explored efficiently in a symbolic way, e.g., using bounded model checking~\cite{BuL11}.
We consider a decomposition in the continuous state space.
Schupp et al.\ perform such a decomposition by syntactic independence~\cite{SchuppNA17}, which corresponds to dynamics with block-diagonal matrices (whereas our decomposition is generally applicable).

For purely continuous systems there exist various decomposition approaches.
In this work we build on~\cite{BogomolovFFVPS18} for LTI systems, which decomposes the system into blocks and exploits the linear dynamics to avoid the wrapping effect.
Other approaches for LTI systems are based on Krylov subspace approximations~\cite{HanK06}, time-scale decomposition~\cite{Dontchev92,GoncharovaO09}, similarity transformations~\cite{KaynamaO10,KaynamaO11}, projectahedra~\cite{GreenstreetM99,YanG08a}, and sub-polyhedra abstract domains~\cite{SeladjiB13}.
Approaches for nonlinear systems are based on projections with differential inclusions~\cite{AsarinD04}, Hamilton-Jacobi methods~\cite{MitchellT03,ChenHT17}, and hybridization with iterative refinement~\cite{ChenS16}.

\paragraph*{Lazy flowpipe computation}

The support-function representation of convex sets can be used to represent a flowpipe (a sequence of sets that covers the behaviors of a system) symbolically~\cite{GirardLG08lgg}.
Only sets that are of interest, e.g., those that intersect with a linear constraint, need then be approximated~\cite{FrehseR12}.
Using our decomposition approach, we can even avoid the symbolic computation in dimensions that are irrelevant to the intersection.
Our approach is independent of the set representation, so it can also be applied in analyses based on, e.g., zonotopes~\cite{Girard05,GirardLG08zonotopes,AlthoffSB10,AlthoffF16}.
Given a linear switching system with a hyperplanar state-space partition, one approach computes ellipsoidal approximations of the reach set on the partition borders, without computing the full reach set~\cite{HamadehG08}.

\paragraph*{Intersection of convex sets}

Performing intersections in low dimensions allows for efficient computations that are not possible in high dimensions.
For example, checking for emptiness of a polyhedron in constraint representation is a feasibility linear program, which can be solved in weakly polynomial time, but solutions in strongly polynomial time are only known in two dimensions~\cite{HochbaumN94}.

In the context of hybrid-system reachability, computing intersections is a major challenge because it usually requires a conversion from efficient set representations (like zonotopes, support function, or Taylor models) to polytopes and back, often entailing additional approximation.
Below we summarize how other approaches tackle the intersection problem.

A coarse approximation of an intersection of a set \X and another set \Y can be
obtained by only detecting a nonempty intersection (which is generally
easier to do) and then taking the original set $\X$ as
overapproximation~\cite{NedialkovM02}.  In general, the intersection
between a polytope and polyhedral constraints (invariants and guards)
can be computed exactly, but such an approach is not
scalable~\cite{ChutinanK03}.  Girard and Le Guernic consider
hyperplanar constraints where reachable states are either zonotopes,
in which case they work in a two-dimensional
projection~\cite{GirardLG08zonotopes}, or general
polytopes~\cite{LeGuernicG09,LeGuernic09}.  The tool SpaceEx
approximates the intersection of polytopes and general polyhedra using
template directions~\cite{FrehseLGDCRLRGDM2011}.  Frehse and Ray
propose an optimization scheme for the intersection of a compact convex set
\X, represented by its support function, and a polyhedron \Y, and this
scheme is exact if \X is a polytope~\cite{FrehseR12}. The problem of
performing intersections can also be cast in terms of finding
separating
hyperplanes~\cite{DBLP:conf/hybrid/FrehseBGSP15,bogomolov-et-al:tacas-2017}. Althoff
et al.\ approximate zonotopes by parallelotopes before considering the
intersection~\cite{AlthoffSB10}.  For must semantics, Althoff and
Krogh use constant-dynamics approximation and obtain a nonlinear
mapping~\cite{AlthoffK12}.  Under certain conditions, Bak et
al.\ apply a model transformation by replacing guard constraints by
time-triggered constraints, for which intersection is
easy~\cite{BakBA17}.

%% file: preliminaries.tex
\section{Preliminaries}\label{sec:preliminaries}

We introduce some notation.
The real numbers are denoted by \R. The origin in $\R^n$ is written $\norigin{n}$.
Given two vectors $x, y \in \R^n$, their dot product is $\dotp{x}{y} := \sum_{i=1}^n x_i \cdot y_i$.
For $p \geq 1$, the $p$-norm of a matrix $A \in \R^{n \times n}$ is denoted $\pnorm{A}$.
The diameter of a set $\X \subseteq \R^n$ is $\Delta_p(\X) := \sup_{x, y \in \mathcal{X}} \pnorm{x - y}$.
The $n$-dimensional unit ball of the $p$-norm is $\Bpn := \{ x \in \R^n \mid 1 \geq \pnorm{x} \}$.
An $n$-dimensional half-space is the set $\{\dotp{a}{x} \leq b \mid x \in \R^n\}$ parameterized by $a \in \R^n$, $b \in \R$.
A polyhedron is an intersection of finitely many half-spaces, and a polytope is a bounded polyhedron.

Given sets $\X \subseteq \R^n$ and $\Y \subseteq \R^m$, scalar $\lambda \in \R$, matrix $A \in \R^{n \times n}$, and vector $b \in \R^n$, we use the following operations on sets:
scaling $\lambda \X := \{\lambda x \mid x \in \X\}$,
linear map $A \X := \{A x \mid x \in \X\}$, Minkowski sum $\X \oplus \Y := \{ x + y \mid x \in \X \text{ and } y \in \Y \}$ (if $n = m$), affine map $\opmap{(A, b)}{\X} := A \X \oplus \{b\}$, Cartesian product $\X\times \Y := \{(x, y) \mid x \in \X, y \in \Y\}$, intersection $\X \cap \Y := \{z \mid z \in \X, z \in \Y\}$ (if $n = m$), and convex hull $\CH(\X) := \{\lambda \cdot x + (1 - \lambda) \cdot y \mid x, y \in \X, 0 \leq \lambda \leq 1\}$.

Given two sets $\X, \Y \subseteq \R^n$, the Hausdorff distance is
\begin{equation*}
	\hausdorff \bigl(\X, \Y \bigr) := \inf_{\varepsilon \in \R} \left\{ \Y \subseteq \X \oplus \varepsilon\Bpn \text{ and } \X \subseteq \Y \oplus \varepsilon\Bpn \right\}.
\end{equation*}
Let $\convexsets \subseteq 2^{\R^n}$ be the set of $n$-dimensional compact convex sets.
For a nonempty set $\X \in \convexsets$, the support function $\rho_\X : \R^n \to \R$ is defined as
\begin{equation*}
	\rho_\X(d) := \max\limits_{x \in \X} \dotp{d}{x}.
\end{equation*}
The Hausdorff distance of two sets $\X, \Y \in \convexsets$ with $\X \subseteq \Y$ can also be expressed in terms of the support function as
\begin{equation*}
	\hausdorff(\X, \Y) = \max_{\pnorm{d} \leq 1} \rho_{\Y}(d) - \rho_{\X}(d).
\end{equation*}

We use $\bl$ to denote the number of blocks in a partition.
Let $\{\pi_j\}_{j=1}^\bl$ be a set of (contiguous) projection matrices that partition a vector $x \in \R^n$ into $x = [\pi_1 x, \ldots, \pi_\bl x]$.
Given a set $\X$ and projection matrices $\{\pi_j\}_{j=1}^\bl$, we call $\pi_j \X$ a \emph{block} of \X and $\bigtimes_j \pi_j \X = \pi_1 \X \times \cdots \times \pi_\bl \X$ the \emph{Cartesian decomposition} with the block structure induced by projections $\pi_j$.
We write $\Xhat$ to indicate a decomposed set (i.e., a Cartesian product of lower-dimensional sets).
For instance, given a nonempty set $\X \in \convexsets$, its box approximation is the Cartesian decomposition into intervals (i.e., one-dimensional blocks).
We can bound the approximation error by the radius of \X.

\begin{proposition}\label{prop:error_decomposition}
	Let $\X \in \convexsets$ be nonempty, $p = \infty$, $r_{\X}^p$ be the radius of the box approximation of \X, and let $\pi_j$ be appropriate projection matrices.
	Then $\hausdorff(\X, \bigtimes_j \pi_j \X) \leq \pnorm{r_{\X}^p}$.
\end{proposition}

\subsection{LTI systems}

An $n$-dimensional \emph{LTI system} $(A, B, \UU)$, with matrices $A \in \R^{n \times n}$, $B \in \R^{n \times m}$, and input domain $\UU \in \convexsets[m]$, is a system of ODEs of the form
\begin{equation}\label{eq:lti}
	\dot{x}(t) = A x(t) + B u(t), \quad u(t) \in \UU.
\end{equation}

We denote the set of all $n$-dimensional LTI systems by $\ltisys$.
From now on, a vector $x \in \R^n$ is also called a (continuous) state.
Given an initial state $x_0 \in \R^n$ and an input signal $u$ such that $u(t) \in \UU$ for all $t$, a \emph{trajectory} of~\eqref{eq:lti} is the unique solution $\traj_{x_0,u}: \Rnonneg \rightarrow \R^n$ with
\begin{equation*}\label{eq:pwa_analytic}
	\traj_{x_0,u}(t) = e^{At} x_0 + \int_0^t e^{A(t-s)} B u(s) \, ds.
\end{equation*}

Given an LTI system $(A, B, \UU)$ and a set $\X_0 \in \convexsets$ of initial states, the \emph{continuous-post operator}, \dwell, computes the set of reachable states for all input signals $u$ over \UU:
\begin{align*}
	& \dwell((A, B, \UU), \X_0) := \\
	& \hspace*{20mm} \{\traj_{x_0, u}(t) \mid t \geq 0, x_0 \in \X_0, u(s) \in \UU \text{ for all } s\}.
\end{align*}

\subsection{Linear hybrid systems}

We briefly introduce the syntax of linear hybrid systems used in this work and refer to the literature for the semantics~\cite{AlurCHH92,Henzinger96}.
An $n$-dimensional linear hybrid system is a tuple $\H = (\var, \locs, \flow, \inv, \grd, \asgn)$ with variables $\var = \{x_1, \dots, x_n\}$, a finite set of locations $\locs$, two functions $\flow: \locs \to \ltisys$ and $\inv: \locs \to \convexsets$ that respectively assign continuous dynamics and an invariant to each location, and two functions $\grd: \locs \times \locs \to \convexsets$ and $\asgn: \locs \times \locs \to \R^{n \times n} \times \R^n$ that respectively assign a guard and an assignment in the form of a deterministic affine map to each pair of locations.
If $\grd(\loc, \loc') \neq \emptyset$, we call $(\loc, \loc')$ a (discrete) transition.

Let $\H = (\var, \locs, \flow, \inv, \grd, \asgn)$ be a linear hybrid system.
A (symbolic) state of \H is a pair $\sstate{\loc}{\X} \in \sstates$.
The \emph{discrete-post operator}, \jump, maps a symbolic state to a set of symbolic states by means of discrete transitions:
\begin{align}
	\jump\sstate{\loc}{\X} := \hspace*{-1mm} \bigcup\limits_{\loc' \in \locs} \hspace*{-1mm} \{ & \sstate{\loc'}{\opcap{\opmap{\asgn(\loc, \loc')}{\phantom{.}} \label{eq:jump} \\[-2mm]
	& {(\opcap{\opcap{\X}{\inv(\loc)}}{\grd(\loc, \loc')})}}{\inv(\loc')}} \} \notag
\end{align}

The \emph{reach set} of \H from a set of initial symbolic states $\somesstates_0$ of \H is the smallest set \somesstates of symbolic states such that
\begin{align}\label{eq:reach}
	\somesstates_0 \cup \bigcup\limits_{\sstate{\loc}{\X} \in \somesstates} \jump\sstate{\loc}{\dwell(\flow(\loc), \X)} \subseteq \somesstates.
\end{align}

We note that the usual definition of the reach set of a hybrid system is stricter wrt.\ invariants.
However, practical analysis tools use the above definition for efficiency reasons.

\medskip

\emph{Safety properties} can be given as a set of symbolic error states that should be avoided and be encoded as the guard of a transition to a new error location.
In our implementation we can also check inclusion in the safe states (the complement) and do not require an encoding with additional transitions.

%% file: algorithm_naive.tex
\section{Reachability analysis of linear hybrid systems}\label{sec:reach_naive}

Our reachability algorithm for linear hybrid systems is centered around the algorithm from~\cite{BogomolovFFVPS18}, called \dwelldeco for convenience, which implements \dwell for LTI systems in a compositional way.
In this section, we first recall the \dwelldeco algorithm.
Two important properties of \dwelldeco are that (1)~the output is a sequence of \emph{decomposed} sets and that (2)~this sequence is computed efficiently in low dimensions.

After explaining the algorithm \dwelldeco, we incorporate it in a standard reachability algorithm for linear hybrid systems.
However, this standard reachability algorithm will not make use of the above-mentioned properties.
This motivates our new algorithm (presented in Section~\ref{sec:reach_new}), which is a modification of both this standard reachability algorithm and \dwelldeco to make optimal use of these properties.

\subsection{Decomposed reachability analysis of LTI systems}\label{sec:reach_lti}

The decomposition-based approach~\cite{BogomolovFFVPS18} follows a flowpipe-construction scheme using time discretization, which we shortly recall here.
Given an LTI system $(A, B, \UU)$ and a set of initial (continuous) states $\X_0$, by fixing a time step $\delta$ we first compute a set $\X(0)$ that overapproximates the reach set up to time $\delta$, a matrix $\Phi = e^{A \delta}$ that captures the dynamics of duration $\delta$, and a set $\V$ that overapproximates the effect of the inputs up to time $\delta$.
We obtain an overapproximation of the reach set in time interval $[k\delta, (k+1)\delta]$, for step $k > 0$, with
\begin{equation*}
	\X(k) := \Phi \X(k-1) \oplus \V = \Phi^k \X(0) \oplus \bigoplusdm_{j = 0}^{k-1} \Phi^j \V.
\end{equation*}

Algorithm \dwelldeco decomposes this scheme.
Fixing some block structure, let $\Xhat(0) := \X_1(0) \times \dots \times \X_\bl(0)$ be the corresponding Cartesian decomposition of $\X(0)$.
We compute a sequence $\Xhat(k) := \X_1(k) \times \dots \times \X_\bl(k)$ such that $\X(k) \subseteq \Xhat(k)$ for every $k$.
Each low-dimensional set $\X_i(k)$ is computed as
\begin{equation*}
	\X_i(k) := \bigoplusdm_{j=1}^\bl (\Phi^k)_{ij} \X_j(0) \oplus \bigoplusdm_{j = 0}^{k-1} [(\Phi^j)_{i1} \cdots (\Phi^j)_{i\bl}] \V.
\end{equation*}
The above sequences $\X(k)$ resp.\ $\Xhat(k)$ are called flowpipes.

\paragraph*{Example}

\begin{figure}[t]
	\centering
	\includegraphics[height=32mm,width=\textwidth,keepaspectratio]{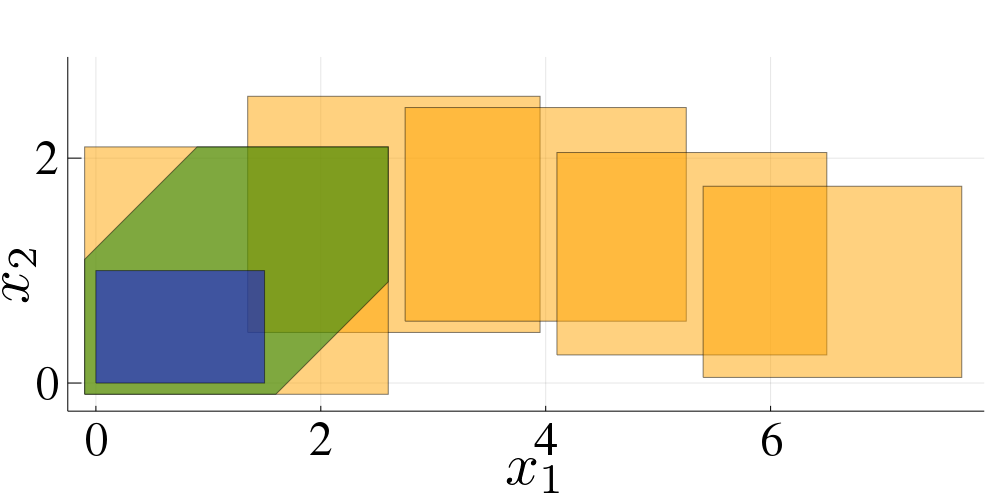}
	\caption{Starting from the set of initial states $\X_0$ (blue set), we first compute the set $\X(0)$ by time discretization (green set), then decompose the set into intervals and obtain $\Xhat(0)$ (orange box around $\X(0)$), and finally compute the (decomposed) flowpipe $\Xhat(1), \dots, \Xhat(4)$ by propagating each of the intervals (other orange sets).}
	\label{fig:postC1}
\end{figure}

\begin{figure}[t]
	\centering
	\includegraphics[height=32mm,width=\textwidth,keepaspectratio]{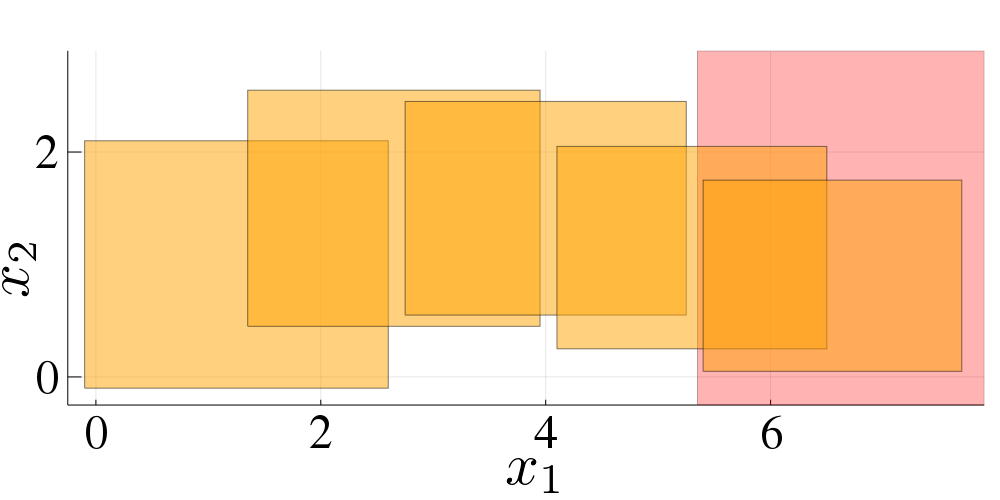}
	\caption{The flowpipe from Figure~\ref{fig:postC1} together with a guard (red).}
	\label{fig:intersection}
\end{figure}

We illustrate the algorithms with a running example throughout the paper.
For illustration purposes, the example is two-dimensional (and hence we decompose into one-dimensional blocks, but we emphasize that the approach also generalizes to higher-dimensional decomposition) and we consider a hybrid system with only a single location and one transition (a self-loop).
Figure~\ref{fig:postC1} depicts the flowpipe construction for the example.

\subsection{Reachability analysis of linear hybrid systems}\label{sec:reach_hybrid}

We now discuss a standard reachability algorithm for hybrid systems.
Essentially, this algorithm interleaves the operators \dwell and \jump following~\eqref{eq:reach} until it finds a fixpoint.
Here we use \dwelldeco as the continuous-post operator.

We first compute a flowpipe $\Xhat = \Xhat(0), \dots, \Xhat(N)$ using \dwelldeco as described above.
Then we use \jump to take a discrete transition.
According to~\eqref{eq:jump}, we want to compute $\opcap{(\opmap{(A, b)}{(\opcap{\opcap{\Xhat}{\set{I}_1}}{\GG})})}{\set{I}_2}$, where $(A, b)$ is a deterministic affine map $\set{I}_1$ and $\set{I}_2$ are the source and target invariant, and $\GG$ is the guard.
Frehse and Ray showed that for such maps the term can be simplified to
\begin{equation}\label{eq:jump_simplified}
	\opmap{(A, b)}{(\opcap{\Xhat}{\GG^*})}
\end{equation}
where the set $\GG^*$ can be statically precomputed~\cite{FrehseR12}, which is usually easy because the sets $\set{I}_1$, \GG, and $\set{I}_2$ are given as polyhedra in constraint representation (also called H-representation, as opposed to the (vertex) V-representation).
Hence we ignore invariants in the rest of the presentation for simplicity.

\paragraph*{Example}

\begin{figure}[t]
	\centering
	\includegraphics[height=32mm,width=\textwidth,keepaspectratio]{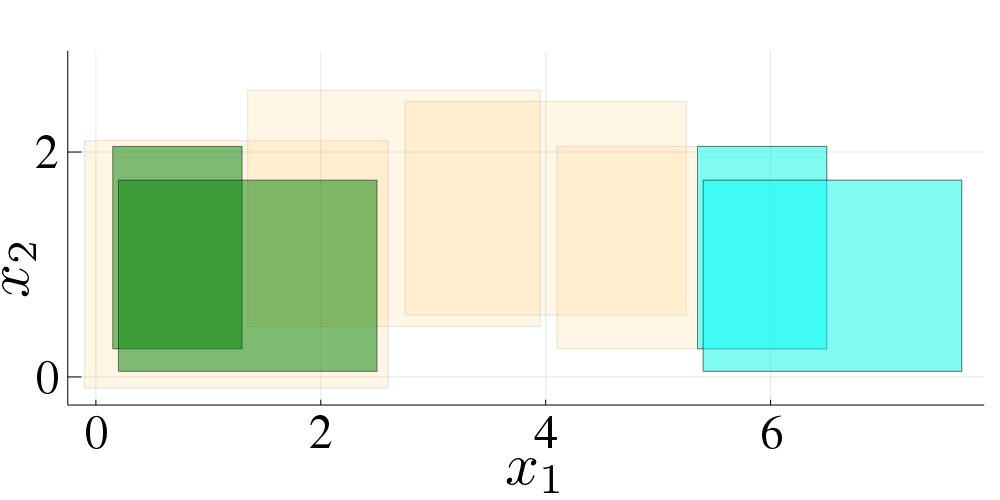}
	\caption{The assignment shifts the intersection of the flowpipe and the guard from Figure~\ref{fig:intersection} (cyan) to the green sets, which are both contained in the first set of the original flowpipe.}
	\label{fig:intersection_assignment_inclusion}
\end{figure}

\begin{figure}[t]
	\centering
	\includegraphics[height=32mm,width=\textwidth,keepaspectratio]{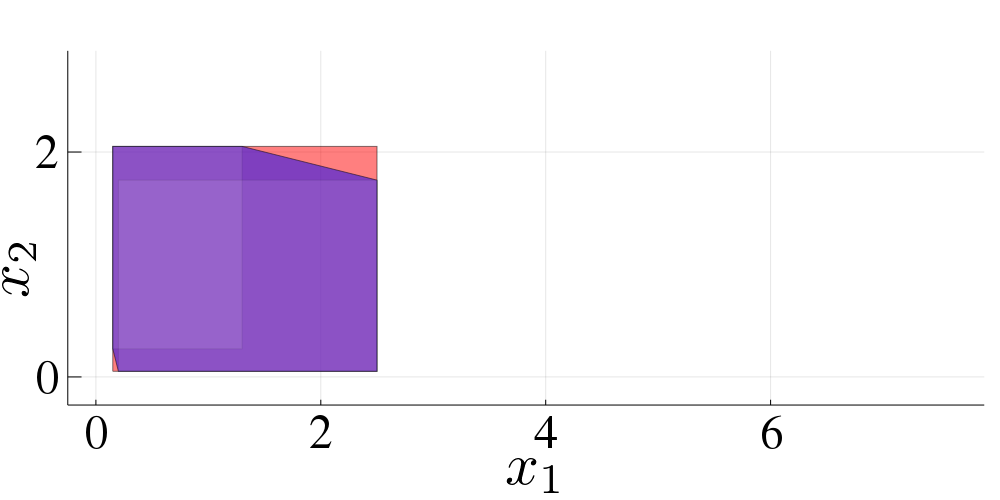}
	\caption{Approximation of the union of the green sets from Figure~\ref{fig:intersection_assignment_inclusion} using the convex hull (purple) and the decomposed convex hull (red).}
	\label{fig:convex_hulls}
\end{figure}

We continue with the flowpipe from Figure~\ref{fig:intersection}.
The guard \GG is a half-space that is constrained in dimension $x_1$ and unconstrained in dimension $x_2$.
Only the last two sets in the flowpipe intersect with the guard.
The assignment here is a translation in dimension $x_1$.
The resulting intersection, before and after the translation, is depicted in Figure~\ref{fig:intersection_assignment_inclusion}.

\medskip

Finally, the algorithm checks for a fixpoint, i.e., for inclusion of the symbolic states we computed with \jump in previously-seen symbolic states.

\paragraph*{Example}

The green set in Figure~\ref{fig:intersection_assignment_inclusion} shows that the two sets we obtained from \jump are contained in $\Xhat(0)$ computed before.
(Recall that in this example we only consider a single location; hence the inclusion of the sets implies inclusion of the symbolic states.)

\medskip

The steps outlined above describe one iteration of the standard reachability algorithm.
Each symbolic state for which the fixpoint check was negative spawns a new flowpipe.
Since this can lead to a combinatorial explosion, one typically applies a technique called \emph{clustering} (cf.~\cite{FrehseLGDCRLRGDM2011}), where symbolic states are merged after the application of \jump.
Here we consider clustering with a convex hull.

\paragraph*{Example}

Assume that the fixpoint check above was negative for both sets that we tested.
In Figure~\ref{fig:convex_hulls} we show the convex hull of the sets in purple.

\medskip

Up to now, we have seen a standard incorporation of an algorithm for the continuous-post operator \dwell (for which we used \dwelldeco) into a reachability algorithm for hybrid systems.
Observe that \dwell was used as a black box.
Consequently, we could not make use of the properties of the specific algorithm \dwelldeco.
In particular, apart from \dwelldeco, we performed all computations in high dimensions.
In the next section, we describe a new algorithm that instead performs all computations in low dimensions.

%% file: algorithm_new.tex
\section{Decomposed reachability analysis}\label{sec:reach_new}

We now present a new, decomposed reachability algorithm for linear hybrid systems.
The algorithm uses a modified version of \dwelldeco for computing flowpipes and has two major performance improvements over the algorithm seen before.

Recall that \dwelldeco computes flowpipes consisting of decomposed sets.
The first improvement is to exploit the decomposed structure to perform all other operations (intersection, affine map, inclusion check, and convex hull) in low dimensions.

The second improvement is to compute flowpipes in a sparse way.
Roughly speaking, we are only interested in those dimensions of a flowpipe that are relevant to determine an intersection with a guard.
We only need to compute the other dimensions of the flowpipe if we detect such an intersection.

We defer proofs in this section to Appendix~\ref{sec:proofs}.

\medskip

The algorithm starts as before:
given an initial (symbolic) state, we compute $\X(0)$ (discretization) and decompose the set to obtain $\Xhat(0)$.
Next we want to compute a flowpipe, and this is where we deviate from the previous algorithm.

\subsection{Computing a sparse flowpipe}

We modify \dwelldeco in order to control the dimensions of the flowpipe.
Recall that the black-box version of \dwelldeco computed the flowpipe $\Xhat(k) = \X_1(k) \times \dots \times \X_\bl(k)$ for $k = 1, \dots, N$, i.e., in all dimensions.
This is usually not necessary for detecting an intersection with a guard.
We will discuss this formally below, but want to establish some intuition first.
Recall the running example from before.
The guard was only constrained in dimension $x_1$.
This means that the bounds of the sets $\Xhat(k) = \X_1(k) \times \X_2(k)$ in dimension $x_2$ are irrelevant.
Consequently, we do not need to compute the sets $\X_2(k)$ at all (at least for the moment).
We only compute those dimensions of a flowpipe that are necessary to determine intersection with the guards.
Identifying these dimensions and projecting the guards accordingly has to be performed only once per transition and is often just a syntactic procedure.

\input{c_post.tex}

In particular, Algorithm~\ref{algorithm:reach} describes the new implementation of \dwelldeco to compute a sparse flowpipe.
The algorithm starts the same way as the original algorithm in~\cite{BogomolovFFVPS18} with a decomposed set in line~\ref{line:initial_states}
and iteratively computes a sparse flowpipe only for the dimensions of interest (line~\ref{line:reach_sparse}).
However, if we detect an intersection with a guard constraint (line~\ref{line:reach_full}), we compute the full-dimensional flowpipe for the corresponding time frame.
The computation of the inputs $\set{\hat{V}_{\text{tmp}}}$ remains the same as in the original algorithm.

\paragraph*{Example}

\begin{figure}[t]
	\centering
	\begin{subfigure}{\linewidth}
		\centering
		\includegraphics[height=32mm,width=\textwidth,keepaspectratio]{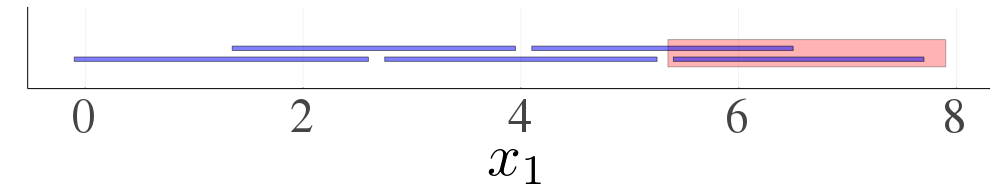}
		\caption{The flowpipe from Figure~\ref{fig:postC1} in dimension $x_1$ only consists of intervals (blue).
	The linear constraint $\GG_1$ (red) is the guard $\GG$ projected to $x_1$.
	For better visibility, we draw the sets thicker and add a slight offset to some of the intervals.}
		\label{fig:postC2_sparse}
	\end{subfigure}
	\\[1mm]
	\begin{subfigure}{\linewidth}
		\centering
		\includegraphics[height=32mm,width=\textwidth,keepaspectratio]{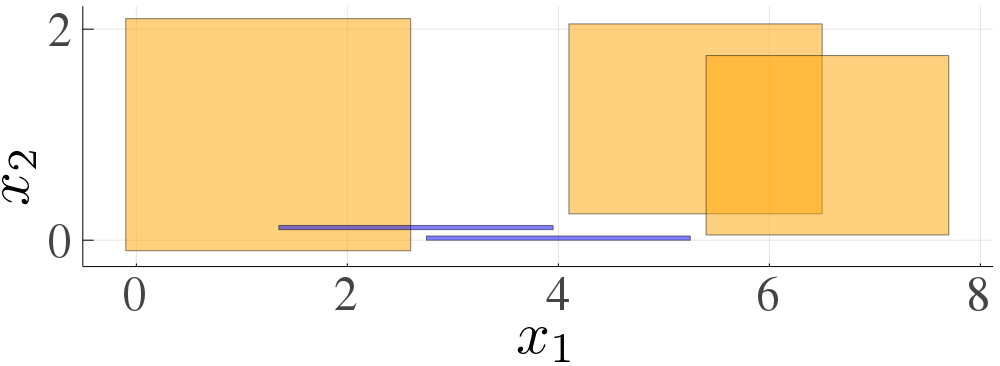}
		\caption{Illustration of the effective flowpipe computation from Figure~\ref{fig:postC1}.
		Only the first set $\Xhat(0)$ and the sets that intersect with the guard ($\Xhat(3)$ and $\Xhat(4)$) are computed in high dimensions.}
		\label{fig:postC2_mixed}
	\end{subfigure}
	\caption{Illustration of the mixed sparse and high-dimensional flowpipe construction.}
	\label{fig:postC2}
\end{figure}

As discussed, we only compute the flowpipe $\X_1(1), \dots, \X_1(4)$ for the first block (in dimension $x_1$), i.e., a sequence of intervals, which is depicted in Figure~\ref{fig:postC2_sparse}.
Projecting the guard to $x_1$, we obtain a ray $\GG_1$.
As expected, we observe an intersection with the guard for the same time steps as before (namely steps $k = 3$ and $k = 4$).

\subsection{Decomposing an intersection}\label{sec:deco_intersection}

Computing the intersection of two $n$-dimensional sets \X and \YG in low dimensions is generally beneficial for performance; yet it is particularly interesting if one of the sets is a polytope that is not represented by its linear constraints.
Common cases are the V-representation or zonotopes represented by their generators, which are used in several approaches~\cite{Girard05,GirardLG08zonotopes,AlthoffSB10,AlthoffF16}.
To compute the (exact) intersection of such a polytope \X with a polyhedron \YG in H-representation, \X needs to be converted to H-representation first.
A polytope with $m$ vertices can have $\Oofl{\binom{m-n/2}{m-n}}$ linear constraints~\cite{McMullen70}.
(For two polytopes in V-representation \emph{in general position} there is a polynomial-time intersection algorithm~\cite{FukudaLL01}, but this assumption is not practical.)
A zonotope with $m$ generators can have $\Oofl{m \binom{m}{n-1}}$ linear constraints~\cite{AlthoffSB10}.
If \YG is a polytope in H-representation, checking disjointness of \X and \YG can also be solved more efficiently in low dimensions, e.g., for $m$ linear constraints in $\Oof{m}$ for $n \leq 3$~\cite{BarbaL15}.

Next we discuss how to apply low-dimensional reasoning to the intersection $\Xhat \cap \YG$ of a decomposed set \Xhat and a polyhedron \YG, or respectively detect emptiness of the intersection $\Xhat \cap \YG$ (which can often be achieved more efficiently).
The key idea is to exploit that \Xhat is decomposed.
For ease of discussion, we consider the case of two blocks (i.e., $\Xhat = \X_1 \times \X_2$).
Below we discuss the two cases that \YG is decomposed or not.

\begin{figure}[t]
	\centering
	\input{intersection_algorithms.tex}
	\vspace*{-8.1cm}
	\caption{Illustration of the different intersection algorithms.
	The task (top row) is to compute the intersection of a decomposed set $\Xhat$ with a guard $\YG$ that is constrained in the blocks 1 and 3 (marked in orange).
	We write $\tilde{\mathcal{X}}$ for the new set after an intersection operation with set $\X$.
	\textbf{a)} The traditional high-dimensional intersection does not make use of the decomposed structure of $\Xhat$.
	\textbf{b)} The low-dimensional intersection projects the guard to each block, computes the block-wise intersection, and embeds the result in high dimensions.
	\textbf{c)} The medium-dimensional intersection projects the guard to the constrained  dimensions, computes the intersection with the Cartesian product of the corresponding sets (here: $\X_1$ and $\X_3$), projects the result to the original block structure, and finally embeds this result in high dimensions.}
	\label{fig:interectoin_algorithms}
\end{figure}
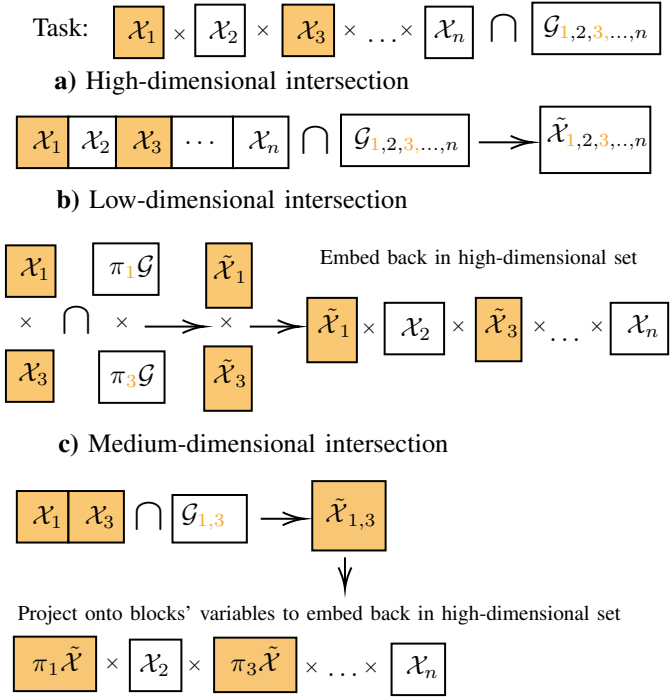

\paragraph*{Intersection of two decomposed sets}

We first consider the case that \YG is also decomposed and agrees with \Xhat on the block structure, i.e., $\YG = \YGhat = \YG_1 \times \YG_2$ and $\X_1, \YG_1 \subseteq \R^{n_1}$ for some $n_1$.
Clearly
\begin{equation*}
	\Xhat \cap \YGhat = (\X_1 \times \X_2) \cap (\YG_1 \times \YG_2) = (\X_1 \cap \YG_1) \times (\X_2 \cap \YG_2)
\end{equation*}
because the Cartesian product and intersection distribute; thus
\begin{equation}\label{eq:decomposed_intersection}
	\Xhat \cap \YGhat = \emptyset \iff (\X_1 \cap \YG_1 = \emptyset) \lor (\X_2 \cap \YG_2 = \emptyset).
\end{equation}

Now consider the second disjunct in~\eqref{eq:decomposed_intersection} and assume that $\YG_2$ is universal.
We get
$
	\X_2 \cap \YG_2 = \emptyset \iff \X_2 = \emptyset.
$
In our context, \Xhat (and hence $\X_2$) is nonempty by construction.
Hence~\eqref{eq:decomposed_intersection} simplifies to
\begin{equation*}\label{eq:decomposed_intersection_one}
	\Xhat \cap \YGhat = \emptyset \iff \X_1 \cap \YG_1 = \emptyset,
\end{equation*}
so we \emph{never} need to compute $\X_2$ to determine whether the intersection is empty.
In practice, the set $\GG^*$ from~\eqref{eq:jump_simplified} takes the role of \YGhat and is often only constrained in \emph{some} dimensions (and hence decomposed and universal in all other dimensions).
We illustrate this algorithm in Figure~\ref{fig:interectoin_algorithms}\,b).

\paragraph*{Intersection of a decomposed and a non-decomposed set}

If \YG is not decomposed in the same block structure as \X, we can still decompose it, at the cost of an approximation error.
Let $\pi_1$ and $\pi_2$ be suitable projection matrices.
Then
\begin{equation*}
	\Xhat \cap \YG \subseteq (\X_1 \cap \pi_1 \YG) \times (\X_2 \cap \pi_2 \YG)
\end{equation*}
and hence
\begin{equation}\label{eq:decomposed_intersection_sufficient}
\begin{split}
	\Xhat \cap \YG = \emptyset & \Longleftarrow (\X_1 \cap \pi_1 \YG = \emptyset) \lor (\X_2 \cap \pi_2 \YG = \emptyset) \\
	&\Longleftarrow \X_1 \cap \pi_1 \YG = \emptyset.
\end{split}
\end{equation}

From~\eqref{eq:decomposed_intersection_sufficient} we obtain 1)~a sufficient test for emptiness of $\Xhat \cap \YG$ in terms of only $\X_1$ and 2)~a more precise sufficient test in terms of $\X_1$ and $\X_2$ in low dimensions.
If both tests fail, we can either fall back to the (exact) test in high dimensions or conservatively assume that the intersection is nonempty.

The precision of the above scheme highly depends on the structure of \Xhat and \YG.
If several (but not all) blocks of \YG are constrained, instead of decomposing \YG into the low-dimensional block structure, one can alternatively compute the intersection for medium-dimensional sets to avoid an approximation error; we apply this strategy in the evaluation (Section~\ref{sec:evaluation}).
If \YG is compact, the following proposition shows that the approximation error is bounded by the maximal entry in the diameters of \Xhat and \YG, and this bound is tight.
This strategy is illustrated in Figure~\ref{fig:interectoin_algorithms}\,c).

\begin{proposition}\label{prop:intersection}
	Let $\Xhat = \bigtimes_j \X_j \in \convexsets$, $\YG \in \convexsets$, $\Xhat \cap \YG \neq \emptyset$, $\YGhat := \bigtimes_j \pi_j \YG$ for appropriate projection matrices $\pi_j$ corresponding to $\X_j$, and $p = \infty$.
	Then
	\begin{equation*}
		\hausdorff(\Xhat \cap \YG, \Xhat \cap \YGhat) \leq \max_j \min(\pnorm{\Delta_p(\X_j)}, \pnorm{\Delta_p(\pi_j \YG)}).
	\end{equation*}
\end{proposition}

\begin{figure}[t]
	\centering
	\includegraphics[height=4cm,width=\textwidth,keepaspectratio]{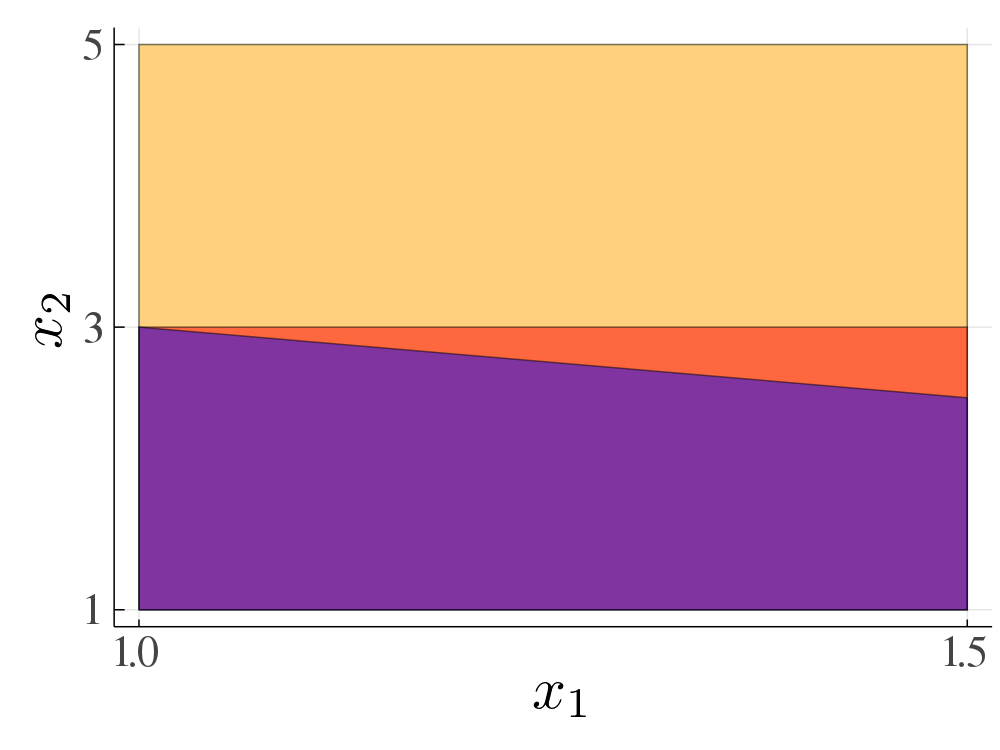}
	\caption{Example of performing the different intersection algorithms.
	 We intersect a three-dimensional hyperrectangle with the ranges $1 \leq x_1 \leq 5$, $1 \leq x_2 \leq 5$, $1 \leq x_3 \leq 5$ and a polyhedral guard with the linear constraints $x_1 + x_2 \leq 4$ and $x_1 \leq 1.5$ (observe that dimension $x_3$ is unconstrained).
	 The exact intersection is the purple set, obtained using the high-dimensional intersection.
	 The red set corresponds to the medium-dimensional intersection.
	 The orange set corresponds to low-dimensional intersection.}
	\label{fig:intersection_example}
\end{figure}

\paragraph*{Example}

Consider Figure~\ref{fig:postC2_mixed}.
Previously we have already identified the intersection with the flowpipe for time steps $k = 3$ and $k = 4$.
The resulting sets are $\Xhat(k) \cap \GG = \X_1(k) \cap \GG_1 \times \X_2(k)$, where $\GG_1$ was the projection of $\GG$ to $x_1$.
We emphasize that we compute the intersections in low dimensions, that we need not compute $\X_2(1)$ and $\X_2(2)$ at all, and that in this example all computations are exact (i.e., we obtain the same sets as in Figure~\ref{fig:intersection_assignment_inclusion}).

Now consider the case that \GG is not decomposed in the same structure as \Xhat, e.g., the hyperplane $x_1 = x_2$.
One option is to decompose \GG to the blocks of \Xhat, i.e., $\GG_1 := \pi_1 \GG$, $\GG_2 := \pi_2 \GG$ and compute the block-wise intersection $(\X_1(k) \cap \GG_1) \times (\X_2(k) \cap \GG_2)$.
Here $\GG_1$ and $\GG_2$ are universal, so we obtain a coarse approximation of the intersection (namely $\Xhat(k)$ itself).
Alternatively, computing the intersection $\Xhat(k) \cap \GG$ is exact but computationally more demanding.
If we assume that the system has higher dimension, e.g., $10$, then computing the intersection in two dimensions (i.e., with a 2D projection $(\X_1(k) \times \X_2(k)) \cap \pi \GG$) is still exact and yet cheaper than computing the intersection in full dimensions (i.e., $\Xhat(k) \cap \GG$).

We exemplify the possible difference between the intersection algorithms in Figure~\ref{fig:intersection_example}.
There we compare the three algorithms visualized in Figure~\ref{fig:interectoin_algorithms}.
The high-dimensional intersection is the most precise as expected because it does not suffer from a projection error, however, the result is a high-dimensional set, instead of a decomposed high-dimensional set.
In addition, this algorithm intersects the sets in three dimensions (including $x_3$) while the other algorithms ignore the unconstrained dimensions.
The medium-dimensional intersection corresponds to the box approximation of the true intersection.
We observe that if we cannot decompose the polyhedron to the same block structure as the decomposed set without a projection error (e.g. projecting $x_1 + x_2 \leq 4$ onto $x_1$ and $x_2$ individually results in one-dimensional universes), then the low-dimensional intersection can produce a large approximation error.
Thanks to the second linear constraint $x_1 \leq 1.5$ and the fact that we treat each half-space separately, we are still able to compute a nontrivial intersection.

\subsection{Decomposing an affine map}

The next step after computing the intersection with the guard is the application of the assignment.
We consider an affine map $A \Xhat \oplus \{b\}$ with $A \in \R^{n \times n}$ and $b \in \R^n$.
Affine-map decomposition has already been presented as part of the operator \dwelldeco~\cite{BogomolovFFVPS18}:
\begin{equation*}\label{eq:decomposed_map_general}
	A \Xhat \oplus \{b\} \subseteq \bigtimesdm_i \bigoplusdm_j A_{ij} \X_j \oplus \{b_i\}
\end{equation*}
where $A_{ij}$ is the block in the $i$-th block row and the $j$-th block column.
We recall an error estimation.

\begin{proposition}{\cite[Prop.~3]{BogomolovFFVPS18}}\label{prop:affine_map}
	Let $\X = \bigtimes_{j=1}^\bl \X_j \in \convexsets$ be nonempty, $A \in \R^{n \times n}$, $q_j := \arg \max_i \pnorm{A_{ij}}$ (the index of the block with the largest matrix norm in the \mbox{$j$-th} block column) so that $\alpha_j := \max_{i \neq q_j} \pnorm{A_{ij}}$ is the second largest matrix norm in the $j$-th block column.
	Let $\alpha_\text{max} := \max_j \alpha_j$ and $\Delta_\text{sum} :=\sum_j \Delta_\infty(\X_j)$.
	Then
	\begin{align*}
		& \hausdorff(A \X, \bigtimesdm_i \bigoplusdm_j A_{ij} \X_j) \\
		={}& \max_{\pnorm{d} \leq 1} \sumdm_{i, j} \rho_{\X_j}(A_{ij}^T d_i) - \rho_{\X_j}\left( \sumdm_k A_{kj}^T d_k \right) \\
		\leq{}& (\bl-1) \sumdm_j \alpha_j \Delta_\infty(\X_j) \leq \frac{n}{2} \alpha_\text{max} \Delta_\text{sum}.
	\end{align*}
\end{proposition}

In particular, if only one block per block column of matrix $A$ is nonzero, the approximation is exact~\cite{BogomolovFFVPS18}.
For example, consider a two-block scenario and a block-diagonal matrix $A$, i.e., $A_{12} = A_{21} = 0$. Then
\begin{align*}
	& \begin{pmatrix} A_{11} & 0 \\ 0 & A_{22} \end{pmatrix} \X_1 \times \X_2 \oplus \{b_1\} \times \{b_2\} \\[-1mm]
	& \hspace{2cm} = (A_{11} \X_1 \oplus \{b_1\}) \times (A_{22} \X_2 \oplus \{b_2\}).
\end{align*}

In practice, affine maps with such a structure are unrealistic for the \dwelldeco operator but typical for assignments.
Prominent cases include resets, translations, and scalings, for which $A$ is even diagonal and hence block diagonal for any block structure.

\paragraph*{Example}

Recall that, after computing the intersections, we ended up with the same sets as in Figure~\ref{fig:intersection_assignment_inclusion}.
In our example, the assignment is a translation in dimension $x_1$.
Hence, as mentioned above, the application of the decomposed assignment is also exact.
In particular, the translation only affects $\X_1(k)$ and we obtain the same result as in Figure~\ref{fig:intersection_assignment_inclusion}.

\subsection{Inclusion check for decomposed sets}

Our algorithm is now fully able to take transitions.
Observe that \emph{all} sets ever occurring in scheme~\eqref{eq:reach} using the algorithm are decomposed.
The following proposition gives an exact low-dimensional fixpoint check under this condition.

\begin{proposition}\label{prop:inclusion}
	Let $\Xhat = \bigtimes_j \X_j \in \convexsets, \YGhat = \bigtimes_j \YG_j \in \convexsets$ be nonempty sets with identical block structure.
	Then
	\[
		\Xhat \subseteq \YGhat \iff \bigwedgedm_j \X_j \subseteq \YG_j.
	\]
\end{proposition}

\subsection{Decomposing a convex hull}

As the last part of the algorithm, we decompose the computation of the convex hull.
We exploit that all sets in the same flowpipe share the same block structure.

\begin{proposition}\label{prop:convex_hull}
	Let $\Xhat = \bigtimes_j \X_j \in \convexsets, \YGhat = \bigtimes_j \YG_j \in \convexsets$ be nonempty sets with identical block structure.
	Then
	\[
		\CH(\Xhat \cup \YGhat) \subseteq \bigtimesdm_j \CH(\X_j \cup \YG_j).
	\]
\end{proposition}

For the decomposition operations proposed before (intersection, affine map, and inclusion), there are common cases where the approximations were exact.
In these cases it is always beneficial to perform the decomposed operations instead of the high-dimensional counterparts.
The decomposition of the convex hull, however, always incurs an approximation error, which we can bound by the radius of the box approximation and by the block-wise difference in bounds.

\begin{proposition}\label{prop:convex_hull_error}
	Let $\Xhat = \bigtimes_j \X_j \in \convexsets, \YGhat = \bigtimes_j \YG_j \in \convexsets$ be nonempty sets with identical block structure and
	let $r^\infty$ be the radius of the box approximation of $\CH(\Xhat \cup \YGhat)$.
	Then
	\begin{align*}
		& \hausdorff(\CH(\Xhat \cup \YGhat), \bigtimesdm_j \CH(\X_j \cup \YG_j)) \\
		\leq{}& \min\left(\!
% 		\frac{n}{2} \Delta_\text{sum},
		\pnorm[\infty]{r^\infty},
		\max_{\pnorm{d} \leq 1} \sumdm_j |\rho_{\X_j}(d_j) - \rho_{\YG_j}(d_j)| \!\right)\!\!.
	\end{align*}
\end{proposition}

\paragraph*{Example}

Figure~\ref{fig:convex_hulls} shows the decomposed convex hull of the sets $\Xhat(3) \cap \GG$ and $\Xhat(4) \cap \GG$ after applying the translation.
Since each block is one-dimensional in our example, we obtain the box approximation.

\subsection{Discussion}\label{sec:discussion}

The combination of the decomposed set operations outlined above results in a reachability algorithm that is sound.
This immediately follows from the fact that all operations compute an overapproximation.

\begin{proposition}[Soundness]
	The analysis using Algorithm~\ref{algorithm:reach} and the aforementioned procedures to apply the decomposed set operations (intersection, affine map, inclusion check, and convex hull) results in an overapproximation of the reach set.
\end{proposition}

\medskip

When using the above algorithm, the question about the choice of the decomposition arises.
Unfortunately, there is generally no automatic way to find the ``best'' decomposition.

Regarding LTI systems, different decompositions generally result in different flowpipes~\cite{BogomolovFFVPS18}.
Unless the LTI system consists of fully decoupled sub-components (which is usually not the case), the only ``best'' decomposition is the degenerate case of a single big block.

Regarding hybrid systems given a fixed decomposition of the LTI systems, there is indeed a unique ``best'' decomposition: a single block for all constrained dimensions.
Finding this ``best'' decomposition is trivial.
However, this block may be too big in terms of computational cost.

We can generally say that a finer partition in the decomposition (i.e., using more and smaller blocks) only ever degrades precision.
However, while lower-dimensional set operations have lower computational cost, a decrease in precision may actually affect the rest of the algorithm in a negative way.
For instance, we may not be able to prove that a transition cannot be taken.
Thus statements about computational cost are generally not possible in the case of hybrid systems.

In the next section we shall see that in practical use cases the decomposition is always beneficial and that using the block size of the constrained dimensions is a good heuristic.

\medskip

Our reachability algorithm benefits greatly from a low number of constrained dimensions, especially in high-dimensional models.
High-dimensional models are scarcely available, but those models that we found supported the hypothesis that this number is typically low.
In particular, the ARCH-COMP competition~\cite{FrehseLGDCRLRGDM2011} represents the state of the art in reachability analysis and considers a scalable model of a powertrain from~\cite{AlthoffK12} where only one dimension is constrained.
In the next section we consider another scalable model with just three constrained dimensions.

%% file: c_post.tex
\begin{algorithm}[t]\footnotesize
	\caption{Function \reach.}
	\label{algorithm:reach}
	\KwIn{%
	$\discsys = (\Phi, \V(\cdot), \X(0))$: discrete system \\
	$N$: total number of steps \\
	\textit{blocks}: list of constrained block indices\\
	\textit{other\_blocks}: list of unconstrained block indices\\
	\textit{constraints}: linear constraints of the outgoing transitions\\
	}
	\vspace*{1mm}
	\SetKwProg{Fn}{function}{}{end}
	\vspace*{2mm}
	\textit{all\_blocks} $\gets$ \textit{blocks} $ \cup$ \textit{other\_blocks} \;
	$\Xhat(0)$ $\gets$ $decompose(\X(0), all\_blocks)$\; \label{line:initial_states}
	$P$ $\gets$ $identity(dim(\Phi))$\;
	$Q$ $\gets$ $\Phi$\;
	$\set{\hat{V}_{\text{tmp}}}$ $\gets$ []\;
	\For{$b_i \in$ \textit{blocks}}{
		$\set{\hat{V}_{\text{tmp}}}[b_i]$ $\gets$ $\{\norigin{dim(b_i)}\}$\;
	}
	\For{$k = 1$ \KwTo $N-1$}{ \label{line:reach_main_loop}
		$\Xhat_{\text{tmp}}$ $\gets$ []\;
		\For{$b_j \in$ \textit{all\_blocks}}{
			$\set{\hat{V}_{\text{tmp}}}[b_i]$ $\gets$ $\set{\hat{V}_{\text{tmp}}}[b_i] \oplus P[b_i, :] \odot \V$\; \label{line:reach_inputs}
		}
		$\Xhat_{\text{tmp}} =$ \textit{reach\_blocks}($\Xhat(0)$, $\Xhat_{\text{tmp}}$, $Q$, $\set{\hat{V}_{\text{tmp}}}$, \textit{blocks}, \textit{all\_blocks})\; \label{line:reach_sparse}
		\If{!isdisjoint($\Xhat_{\text{tmp}}$, constraints)}{\label{line:reach_full}
			\tcp{compute high-dimensional sets only if intersection with guards is nonempty}
			$\Xhat_{\text{tmp}} =$ \textit{reach\_blocks}($\Xhat(0)$, $\Xhat_{\text{tmp}}$, $Q$, $\set{\hat{V}_{\text{tmp}}}$, \textit{other\_blocks}, \textit{all\_blocks})\;
		}
		$\Xhat(k)$ $\gets$ $\Xhat_{\text{tmp}}$\; \label{line:reach_result}
		$P$ $\gets$ $Q$\;
		$Q$ $\gets$ $Q \cdot \Phi$\;
	}
	\textit{return} $[\Xhat(0), \Xhat(1), ..., \Xhat(N-1)]$\;
	\BlankLine
	\Fn{reach\_blocks($\Xhat(0)$, $\Xhat_{\text{tmp}}$, $Q$, $\set{\hat{V}_{\text{tmp}}}$, \textit{blocks}, \textit{all\_blocks})}{
		\For{$b_i \in$ \textit{blocks}}{ \label{line:reach_computation_start}
				$\Xhat_{\text{tmp}}[b_i]$ $\gets$ $\{\norigin{dim(b_i)} \}$\;
				\For{$b_j \in$ \textit{all\_blocks}}{
				$\Xhat_{\text{tmp}}[b_i]$ $\gets$ $\Xhat_{\text{tmp}}[b_i] \oplus Q[b_i, b_j] \odot \Xhat(0)[b_j]$\label{line:reach_innermost}\;
			}
			$\Xhat_{\text{tmp}}[b_i]$ $\gets$ $\Xhat_{\text{tmp}}[b_i] \oplus \set{\hat{V}_{\text{tmp}}}[b_i]$\; \label{line:reach_states_plus_inputs}
		} \label{line:reach_computation_end}
		\textit{return} $\Xhat_{\text{tmp}}$\;
	}
\end{algorithm}

%% file: intersection_algorithms.tex
\tikzset{every picture/.style={line width=0.75pt}} %set default line width to 0.75pt

\begin{tikzpicture}[x=0.75pt,y=0.75pt,yscale=-1,xscale=1]
%uncomment if require: \path (0,361); %set diagram left start at 0, and has height of 361

%Shape: Rectangle [id:dp2553008931357079]
\draw  [fill={rgb, 255:red, 248; green, 166; blue, 28 }  ,fill opacity=0.62 ] (186.5,8) -- (211.5,8) -- (211.5,34) -- (186.5,34) -- cycle ;
%Shape: Rectangle [id:dp5922779710070258]
\draw  [fill={rgb, 255:red, 255; green, 255; blue, 255 }  ,fill opacity=1 ] (227.5,7) -- (252.5,7) -- (252.5,33) -- (227.5,33) -- cycle ;
%Shape: Rectangle [id:dp992156525712647]
\draw  [fill={rgb, 255:red, 255; green, 255; blue, 255 }  ,fill opacity=1 ] (343.5,8) -- (367.5,8) -- (367.5,34) -- (343.5,34) -- cycle ;
%Shape: Rectangle [id:dp9296709930579334]
\draw   (216,254.5) -- (253.5,254.5) -- (253.5,278.5) -- (216,278.5) -- cycle ;
%Shape: Rectangle [id:dp8444374539796438]
\draw  [fill={rgb, 255:red, 245; green, 166; blue, 35 }  ,fill opacity=0.62 ] (137.5,63.5) -- (163.5,63.5) -- (163.5,89.5) -- (137.5,89.5) -- cycle ;
%Shape: Rectangle [id:dp20769465014841337]
\draw  [fill={rgb, 255:red, 255; green, 255; blue, 255 }  ,fill opacity=1 ] (163.5,63.5) -- (187.5,63.5) -- (187.5,89.5) -- (163.5,89.5) -- cycle ;
%Shape: Rectangle [id:dp40342671156571464]
\draw  [fill={rgb, 255:red, 255; green, 255; blue, 255 }  ,fill opacity=1 ] (246.5,63.5) -- (274.5,63.5) -- (274.5,89.5) -- (246.5,89.5) -- cycle ;
%Shape: Rectangle [id:dp7093433029843932]
\draw   (301,62.5) -- (365.5,62.5) -- (365.5,88.5) -- (301,88.5) -- cycle ;
%Shape: Rectangle [id:dp42676290849568144]
\draw  [fill={rgb, 255:red, 255; green, 255; blue, 255 }  ,fill opacity=1 ] (215.5,63.5) -- (246.5,63.5) -- (246.5,89.5) -- (215.5,89.5) -- cycle ;
%Shape: Rectangle [id:dp42969759872191315]
\draw  [fill={rgb, 255:red, 248; green, 166; blue, 28 }  ,fill opacity=0.62 ] (137.5,252.5) -- (163.5,252.5) -- (163.5,278.5) -- (137.5,278.5) -- cycle ;
%Shape: Rectangle [id:dp7863592850302028]
\draw  [fill={rgb, 255:red, 248; green, 166; blue, 28 }  ,fill opacity=0.62 ] (163.5,252.5) -- (191.5,252.5) -- (191.5,278.5) -- (163.5,278.5) -- cycle ;
%Shape: Rectangle [id:dp24547691345240952]
\draw  [fill={rgb, 255:red, 248; green, 166; blue, 28 }  ,fill opacity=0.62 ] (132,128.5) -- (157,128.5) -- (157,154.5) -- (132,154.5) -- cycle ;
%Shape: Rectangle [id:dp7785988081222897]
\draw  [fill={rgb, 255:red, 248; green, 166; blue, 28 }  ,fill opacity=0.62 ] (132,181.5) -- (156,181.5) -- (156,207.5) -- (132,207.5) -- cycle ;
%Shape: Rectangle [id:dp7138070237008465]
\draw   (176,127.5) -- (209,127.5) -- (209,153.5) -- (176,153.5) -- cycle ;
%Shape: Rectangle [id:dp22520640878308518]
\draw   (401.5,58.5) -- (457,58.5) -- (457,92) -- (401.5,92) -- cycle ;
%Shape: Rectangle [id:dp1329856724772811]
\draw  [fill={rgb, 255:red, 248; green, 166; blue, 28 }  ,fill opacity=0.62 ] (233,127.5) -- (256,127.5) -- (256,160) -- (233,160) -- cycle ;
%Shape: Rectangle [id:dp20396020245022073]
\draw  [fill={rgb, 255:red, 248; green, 166; blue, 28 }  ,fill opacity=0.62 ] (232,179.5) -- (256,179.5) -- (256,212) -- (232,212) -- cycle ;
%Straight Lines [id:da2939761120358828]
\draw    (303,286) -- (303,300) ;
\draw [shift={(303,302)}, rotate = 270] [color={rgb, 255:red, 0; green, 0; blue, 0 }  ][line width=0.75]    (10.93,-3.29) .. controls (6.95,-1.4) and (3.31,-0.3) .. (0,0) .. controls (3.31,0.3) and (6.95,1.4) .. (10.93,3.29)   ;
%Shape: Rectangle [id:dp2921785366355687]
\draw  [fill={rgb, 255:red, 255; green, 255; blue, 255 }  ,fill opacity=1 ] (194.5,327.5) -- (220.5,327.5) -- (220.5,353.5) -- (194.5,353.5) -- cycle ;
%Shape: Rectangle [id:dp14306916537700176]
\draw  [fill={rgb, 255:red, 255; green, 255; blue, 255 }  ,fill opacity=1 ] (326.5,329.5) -- (353.5,329.5) -- (353.5,355.5) -- (326.5,355.5) -- cycle ;
%Shape: Rectangle [id:dp509985904029991]
\draw   (323,157.5) -- (353,157.5) -- (353,183.5) -- (323,183.5) -- cycle ;
%Shape: Rectangle [id:dp9854290548101061]
\draw  [fill={rgb, 255:red, 255; green, 255; blue, 255 }  ,fill opacity=1 ] (437,157.5) -- (466,157.5) -- (466,183.5) -- (437,183.5) -- cycle ;
%Straight Lines [id:da594024050447282]
\draw    (255,169.5) -- (278,169.5) ;
\draw [shift={(280,169.5)}, rotate = 180] [color={rgb, 255:red, 0; green, 0; blue, 0 }  ][line width=0.75]    (10.93,-3.29) .. controls (6.95,-1.4) and (3.31,-0.3) .. (0,0) .. controls (3.31,0.3) and (6.95,1.4) .. (10.93,3.29)   ;
%Shape: Rectangle [id:dp7238374785988126]
\draw  [fill={rgb, 255:red, 248; green, 166; blue, 28 }  ,fill opacity=0.62 ] (271.5,7) -- (297.5,7) -- (297.5,33) -- (271.5,33) -- cycle ;
%Shape: Rectangle [id:dp5653586023979327]
\draw  [fill={rgb, 255:red, 248; green, 166; blue, 28 }  ,fill opacity=0.62 ] (187.5,63.5) -- (215.5,63.5) -- (215.5,89.5) -- (187.5,89.5) -- cycle ;
%Straight Lines [id:da4957333016760306]
\draw    (260.5,266.5) -- (280.5,266.5) ;
\draw [shift={(282.5,266.5)}, rotate = 180] [color={rgb, 255:red, 0; green, 0; blue, 0 }  ][line width=0.75]    (10.93,-3.29) .. controls (6.95,-1.4) and (3.31,-0.3) .. (0,0) .. controls (3.31,0.3) and (6.95,1.4) .. (10.93,3.29)   ;
%Straight Lines [id:da47752971408070133]
\draw    (370.5,76.5) -- (395.5,76.5) ;
\draw [shift={(397.5,76.5)}, rotate = 180] [color={rgb, 255:red, 0; green, 0; blue, 0 }  ][line width=0.75]    (10.93,-3.29) .. controls (6.95,-1.4) and (3.31,-0.3) .. (0,0) .. controls (3.31,0.3) and (6.95,1.4) .. (10.93,3.29)   ;
%Straight Lines [id:da059394365124673776]
\draw    (202,169.5) -- (230,169.5) ;
\draw [shift={(232,169.5)}, rotate = 180] [color={rgb, 255:red, 0; green, 0; blue, 0 }  ][line width=0.75]    (10.93,-3.29) .. controls (6.95,-1.4) and (3.31,-0.3) .. (0,0) .. controls (3.31,0.3) and (6.95,1.4) .. (10.93,3.29)   ;
%Shape: Rectangle [id:dp106294325664261]
\draw   (397.5,6) -- (462,6) -- (462,32) -- (397.5,32) -- cycle ;
%Shape: Rectangle [id:dp09280169185458065]
\draw  [fill={rgb, 255:red, 245; green, 166; blue, 35 }  ,fill opacity=0.62 ] (286.5,247.5) -- (323.5,247.5) -- (323.5,282.5) -- (286.5,282.5) -- cycle ;
%Shape: Rectangle [id:dp7862120020571837]
\draw  [fill={rgb, 255:red, 248; green, 166; blue, 28 }  ,fill opacity=0.62 ] (284,155) -- (309,155) -- (309,187.5) -- (284,187.5) -- cycle ;
%Shape: Rectangle [id:dp553109981174877]
\draw  [fill={rgb, 255:red, 248; green, 166; blue, 28 }  ,fill opacity=0.62 ] (369,155) -- (392,155) -- (392,187) -- (369,187) -- cycle ;
%Shape: Rectangle [id:dp4741617728707652]
\draw  [fill={rgb, 255:red, 245; green, 166; blue, 35 }  ,fill opacity=0.62 ] (237,326.5) -- (278.5,326.5) -- (278.5,354) -- (237,354) -- cycle ;
%Shape: Rectangle [id:dp7896777805770767]
\draw  [fill={rgb, 255:red, 245; green, 166; blue, 35 }  ,fill opacity=0.62 ] (136,326) -- (177.5,326) -- (177.5,353.5) -- (136,353.5) -- cycle ;
%Shape: Rectangle [id:dp8839478232288533]
\draw   (179,182) -- (212,182) -- (212,208) -- (179,208) -- cycle ;

% Text Node
\draw (626,656) node   [align=left] {$ $};
% Text Node
\draw (201,21) node   [align=left] {$\displaystyle \mathcal{X}_{1}$};
% Text Node
\draw (348,17) node   [align=left] {$ $};
% Text Node
\draw (323,25) node   [align=left] {$\displaystyle \dotsc $};
% Text Node
\draw (153,76.5) node   [align=left] {$\displaystyle \mathcal{X}_{1}$};
% Text Node
\draw (177.5,76.5) node   [align=left] {$\displaystyle \mathcal{X}_{2}$};
% Text Node
\draw (262.5,76.5) node   [align=left] {$\displaystyle \mathcal{X}_{n}$};
% Text Node
\draw (231,76.5) node   [align=left] {$\displaystyle \dotsc $};
% Text Node
\draw (153,265.5) node   [align=left] {$\displaystyle \mathcal{X}_{1}$};
% Text Node
\draw (179.5,265.5) node   [align=left] {$\displaystyle \mathcal{X}_{3}$};
% Text Node
\draw (146.75,140.5) node   [align=left] {$\displaystyle \mathcal{X}_{1}$};
% Text Node
\draw (146,193.5) node   [align=left] {$\displaystyle \mathcal{X}_{3}$};
% Text Node
\draw (383,19) node  [font=\LARGE] [align=left] {$\displaystyle \cap $};
% Text Node
\draw (288,75.5) node  [font=\LARGE] [align=left] {$\displaystyle \cap $};
% Text Node
\draw (204,265.5) node  [font=\LARGE] [align=left] {$\displaystyle \cap $};
% Text Node
\draw (167.5,166.5) node  [font=\LARGE] [align=left] {$\displaystyle \cap $};
% Text Node
% \draw (430,48.5) node   [align=left] {{\footnotesize Result set}};
% Text Node
% \draw (305,236.5) node   [align=left] {{\footnotesize Result set}};
% Text Node
\draw (245,140.5) node   [align=left] {$\displaystyle \textcolor[rgb]{0,0,0}{\tilde{\mathcal{X}}}\textcolor[rgb]{0,0,0}{_{\textcolor[rgb]{0,0,0}{1}}}$};
% Text Node
\draw (245,193) node   [align=left] {$\displaystyle \textcolor[rgb]{0,0,0}{\tilde{\mathcal{X}}}\textcolor[rgb]{0,0,0}{_{3}}$};
% Text Node
% \draw (243.5,118.5) node   [align=left] {{\footnotesize Result set}};
% Text Node
\draw (290,315) node  [font=\footnotesize] [align=left] {Project onto blocks' variables to embed back in high-dimensional set};
% Text Node
\draw (158,338) node   [align=left] {$\displaystyle \textcolor[rgb]{0,0,0}{\pi }\textcolor[rgb]{0,0,0}{_{1}}\textcolor[rgb]{0,0,0}{\tilde{\mathcal{X}}}$};
% Text Node
\draw (303,345) node   [align=left] {$\displaystyle \dotsc $};
% Text Node
\draw (338.75,170.5) node   [align=left] {$\displaystyle \mathcal{X}_{2}$};
% Text Node
\draw (415,176) node   [align=left] {$\displaystyle \dotsc $};
% Text Node
\draw (370.5,135.5) node  [font=\footnotesize] [align=left] {Embed back in high-dimensional set};
% Text Node
\draw (203.5,76.5) node   [align=left] {$\displaystyle \mathcal{X}_{3}$};
% Text Node
\draw (214,17) node [anchor=north west][inner sep=0.75pt]  [font=\scriptsize] [align=left] {$\displaystyle {\textstyle \times }$};
% Text Node
\draw (256,15) node [anchor=north west][inner sep=0.75pt]  [font=\scriptsize] [align=left] {$\displaystyle {\textstyle \times }$};
% Text Node
\draw (185,162.5) node [anchor=north west][inner sep=0.75pt]  [font=\scriptsize] [align=left] {$\displaystyle {\textstyle \times }$};
% Text Node
\draw (136.5,162.5) node [anchor=north west][inner sep=0.75pt]  [font=\scriptsize] [align=left] {$\displaystyle {\textstyle \times }$};
% Text Node
\draw (237.5,162.5) node [anchor=north west][inner sep=0.75pt]  [font=\scriptsize] [align=left] {$\displaystyle {\textstyle \times }$};
% Text Node
\draw (180,334.5) node [anchor=north west][inner sep=0.75pt]  [font=\scriptsize] [align=left] {$\displaystyle {\textstyle \times }$};
% Text Node
\draw (221.5,335.5) node [anchor=north west][inner sep=0.75pt]  [font=\scriptsize] [align=left] {$\displaystyle {\textstyle \times }$};
% Text Node
\draw (280.5,336.5) node [anchor=north west][inner sep=0.75pt]  [font=\scriptsize] [align=left] {$\displaystyle {\textstyle \times }$};
% Text Node
\draw (310.5,336.5) node [anchor=north west][inner sep=0.75pt]  [font=\scriptsize] [align=left] {$\displaystyle {\textstyle \times }$};
% Text Node
\draw (300,15) node [anchor=north west][inner sep=0.75pt]  [font=\scriptsize] [align=left] {$\displaystyle {\textstyle \times }$};
% Text Node
\draw (328.5,15) node [anchor=north west][inner sep=0.75pt]  [font=\scriptsize] [align=left] {$\displaystyle {\textstyle \times }$};
% Text Node
\draw (424.5,164.5) node [anchor=north west][inner sep=0.75pt]  [font=\scriptsize] [align=left] {$\displaystyle {\textstyle \times }$};
% Text Node
\draw (395.5,164.5) node [anchor=north west][inner sep=0.75pt]  [font=\scriptsize] [align=left] {$\displaystyle {\textstyle \times }$};
% Text Node
\draw (309.5,165.5) node [anchor=north west][inner sep=0.75pt]  [font=\scriptsize] [align=left] {$\displaystyle {\textstyle \times }$};
% Text Node
\draw (354.5,164.5) node [anchor=north west][inner sep=0.75pt]  [font=\scriptsize] [align=left] {$\displaystyle {\textstyle \times }$};
% Text Node
\draw (242,21) node   [align=left] {$\displaystyle \mathcal{X}_{2}$};
% Text Node
\draw (286.25,21) node   [align=left] {$\displaystyle \mathcal{X}_{3}$};
% Text Node
\draw (356,20.5) node   [align=left] {$\displaystyle \mathcal{X}_{n}$};
% Text Node
\draw (430.5,72) node   [align=left] {$\displaystyle \tilde{\mathcal{X}}_{\textcolor[rgb]{0.96,0.65,0.14}{1}\textcolor[rgb]{0,0,0}{,2,}\textcolor[rgb]{0.96,0.65,0.14}{3}\textcolor[rgb]{0,0,0}{,..} ,n}$};
% Text Node
\draw (429.75,19) node   [align=left] {$\displaystyle \mathcal{G}_{\textcolor[rgb]{0.96,0.65,0.14}{1}\textcolor[rgb]{0,0,0}{,2,}\textcolor[rgb]{0.96,0.65,0.14}{3,} ...,n}$};
% Text Node
\draw (231.75,265.5) node   [align=left] {$ $$\displaystyle \mathcal{G}_{\textcolor[rgb]{0.96,0.65,0.14}{1,3}}$};
% Text Node
\draw (306,264) node   [align=left] {$\displaystyle \textcolor[rgb]{0,0,0}{\tilde{\mathcal{X}}}\textcolor[rgb]{0,0,0}{_{\textcolor[rgb]{0,0,0}{1,3}}}$};
% Text Node
\draw (207,340.5) node   [align=left] {$\displaystyle \mathcal{X}_{2}$};
% Text Node
\draw (342,341.5) node   [align=left] {$\displaystyle \mathcal{X}_{n}$};
% Text Node
\draw (258,338) node   [align=left] {$\displaystyle \textcolor[rgb]{0,0,0}{\pi }\textcolor[rgb]{0,0,0}{_{3}}\textcolor[rgb]{0,0,0}{\tilde{\mathcal{X}}}$};
% Text Node
\draw (195.5,139.5) node   [align=left] {$\displaystyle \textcolor[rgb]{0,0,0}{\pi }\textcolor[rgb]{0.96,0.65,0.14}{_{1}}$$\displaystyle \mathcal{G}$};
% Text Node
\draw (196,194.5) node   [align=left] {$\displaystyle \textcolor[rgb]{0,0,0}{\pi }\textcolor[rgb]{0.96,0.65,0.14}{_{3}}$$\displaystyle \mathcal{G}$};
% Text Node
\draw (297.75,167.5) node   [align=left] {$\displaystyle \textcolor[rgb]{0,0,0}{\tilde{\mathcal{X}}}\textcolor[rgb]{0,0,0}{_{\textcolor[rgb]{0,0,0}{1}}}$};
% Text Node
\draw (382,167.5) node   [align=left] {$\displaystyle \textcolor[rgb]{0,0,0}{\tilde{\mathcal{X}}}\textcolor[rgb]{0,0,0}{_{3}}$};
% Text Node
\draw (453.5,169) node   [align=left] {$\displaystyle \mathcal{X}_{n}$};
% Text Node
\draw (334.25,75.5) node   [align=left] {$\displaystyle \mathcal{G}_{\textcolor[rgb]{0.96,0.65,0.14}{1}\textcolor[rgb]{0,0,0}{,2,}\textcolor[rgb]{0.96,0.65,0.14}{3,} ...,n}$};
% Text Node
\draw (155,39) node [anchor=north west][inner sep=0.75pt]   [align=left] {\textbf{a)} High-dimensional intersection};
% Text Node
\draw (157,222) node [anchor=north west][inner sep=0.75pt]   [align=left] {\textbf{c)} Medium-dimensional intersection};
% Text Node
\draw (156,99) node [anchor=north west][inner sep=0.75pt]   [align=left] {\textbf{b)} Low-dimensional intersection};
% Text Node
\draw (144,12) node [anchor=north west][inner sep=0.75pt]   [align=left] {Task:};

\end{tikzpicture}

%% file: evaluation.tex
\section{Evaluation}\label{sec:evaluation}

We implemented the ideas presented in Section~\ref{sec:reach_new} in JuliaReach~\cite{JuliaReach,BogomolovFFPS19}.
% We are able to process systems with up to 512 dimensions for a reasonable time. We compare our approach with the classic hybrid algorithm, where interesctions, linear maps and other operations are taken in high dimension.
The code is publicly available~\cite{JuliaReach}.
We performed the experiments presented in this section on a Mac notebook with an Intel i5 CPU@3.1\,GHz and 16\,GB RAM.

\subsection{Benchmark descriptions}
We evaluate our implementation on a number of benchmarks taken from the HyPro model library~\cite{HyProBench}, from the ARCH-COMP 2019 competition~\cite{althoff2019arch}, and a scalable model from~\cite{FrehseLGDCRLRGDM2011}.
To demonstrate the qualitative performance of our approach, we verify a safety property for each benchmark, which requires precise approximations in each step of the algorithm.
We briefly describe the benchmarks below.

\textit{Linear switching system}. This five-dimensional model taken from~\cite{HyProBench} is a linear hybrid system with five locations of different controlled, randomly-generated continuous dynamics stabilized by an LQR controller. The discrete structure has a ring topology with transitions determined heuristically from simulations. The safety property for this system is $x_1 > -1.2$.

\textit{Spacecraft rendezvous.} This model with five dimensions represents a spacecraft steering toward a passive target in space~\cite{ChanM17}. We use a linearized version of this model with three locations. We consider two scenarios, one where the spacecraft successfully approaches the target and another one where a mission abort occurs at $t = 120$\,min. For the safety properties we refer to~\cite{althoff2019arch}.

\textit{Platooning.} This ten-dimensional model with two locations represents a platoon of three vehicles with communication loss at deterministic times~\cite{MakhloufK14}.
The safety property enforces a minimum distance $d$ between the vehicles: $\bigwedge_{x \in \{x_1, x_4, x_7\}} x \geq -d$.
We consider both a time-bounded setting with $d = 42$ and a time-unbounded setting with $d = 50$; note that in the latter setting a fixpoint must be found.

\textit{Filtered oscillator.} This scalable model consists of a two-dimensional switched oscillator (dimensions $x$ and $y$) and a parametric number of filters (here: 64--1024) which smooth $x$~\cite{FrehseLGDCRLRGDM2011}. We fixed the maximum number of transitions to five by adding a new variable. The safety property is $y < 0.5$.

\begin{table*}[t]
	\centering
	\begin{threeparttable}
	\caption{
	For each benchmark we report the number of dimension and the number of constrained dimensions (\emph{Dim. (constr.)}), the step size of the time discretization (\emph{Step}) and the run time of the different algorithms as described in Section~\ref{sec:tool_descriptions} (e.g., ``Deco'' refers to the decomposed algorithm presented in this paper) in seconds (where the fastest solution is marked in bold face). For benchmark instances with parenthesized computation time, the safety property could not be proven because the overapproximation was too coarse, in which case the computation terminated as soon as the violation was detected.
	``TO'' and ``OOM''  indicate a timeout of $5 \times 10^4$ seconds and an out-of-memory error, respectively.
	}
	\label{tab:benchmarks}
	\renewcommand{\arraystretch}{1.1}
	\def\tabspaceborder{\hspace*{2mm}}
	\def\tabspace{\hspace*{4.5mm}}
	\def\mcc#1{\multicolumn{1}{c @{\tabspace}}{#1}}
	\def\TO{\hspace*{5mm}TO}
	\def\OOM{\hspace*{3.5mm}OOM}
	\begin{tabular}{@{\tabspaceborder} l @{} r @{\tabspace} l @{\tabspace} l @{\tabspace} l @{\tabspace} l @{\tabspace} l @{\tabspace} l @{\tabspaceborder}}
		\toprule
		Benchmark           & \mcc{Dim. (constr.)} & Step & \hspace*{2mm} Deco   & LazySupp                       & LazyOptim          & SpaceEx LGG                    & SpaceEx STC \\
		\midrule
		linear\_switching   &    5 (1) & 0.0001 & $\mathbf{2.50 \times 10^0}$    & $1.27 \times 10^1$    & $2.81\times 10^1$  & $2.60 \times 10^1$             & $2.30 \times 10^1$ \\
		spacecraft\_noabort &    5 (4) & 0.04   & $5.30 \times 10^0$             & $3.42 \times 10^0$             & $2.19\times 10^2$  & $1.18 \times 10^{0}$          & $\mathbf{3.50 \times 10^{-1}}$ \\
		spacecraft\_120     &    5 (5) & 0.04   & $5.30 \times 10^0$             & $2.10 \times 10^0$             & $4.30 \times 10^1$ & $1.91 \times 10^0$ & $\mathbf{8.10 \times 10^{-1}}$ \\
		platoon\_bounded    &   10 (4) & 0.01   & $\mathbf{1.30 \times 10^{-1}}$ & $1.60 \times 10^{-1}$          & $5.69 \times 10^0$ & $5.55 \times 10^{0}$          & $1.60 \times 10^0$ \\
		platoon\_unbounded  &   10 (4) & 0.03   & $\mathbf{1.08 \times 10^0}$    & $1.16 \times 10^0$             & $4.96 \times 10^1$ & $3.46 \times 10^1$             & $6.50 \times 10^1$ \\
		filtered\_osc64     &   67 (3) & 0.01   & $\mathbf{2.81 \times 10^0}$    & \unverified{$7.43 \times 10^0$} & $5.63\times 10^2$  & $2.04 \times 10^1$             & $3.25 \times 10^1$ \\
		filtered\_osc128    &  131 (3) & 0.01   & $\mathbf{7.95 \times 10^0}$    & \unverified{$4.29 \times 10^1$} & $1.79\times 10^3$  & $1.69 \times 10^2$             & $4.67 \times 10^3$ \\
		filtered\_osc256    &  259 (3) & 0.01   & $\mathbf{2.80 \times 10^1}$    & \unverified{$9.19 \times 10^1$} & $9.99\times 10^4$  & $8.70 \times 10^3$             & \OOM \\
		filtered\_osc512    &  515 (3) & 0.01   & $\mathbf{1.13 \times 10^2}$    & \unverified{$4.73 \times 10^2$} &                \TO &                         \TO    & \TO \\
		filtered\_osc1024   & 1027 (3) & 0.01   & $\mathbf{5.09 \times 10^2}$    & \unverified{$5.11 \times 10^3$} &                \TO &                         \TO    & \TO \\
		\bottomrule
	\end{tabular}
	\begin{tablenotes}
		$^\dag$The safety property could not be proven by this tool because the overapproximation was too coarse.
	\end{tablenotes}
	\end{threeparttable}
	\renewcommand{\arraystretch}{1}
\end{table*}

\subsection{Tool descriptions}\label{sec:tool_descriptions}
We compare our implementation in JuliaReach to two other algorithms available in the same tool. All three algorithms use the same decomposition-based continuous-post operator \dwelldeco~\cite{BogomolovFFPS19}, so the main difference between these algorithms is the intersection operation with discrete jumps, which allows for a direct comparison of the approach presented in this paper.
The existing implementations can be seen as instances of the algorithm in Section~\ref{sec:reach_naive}.
Furthermore, we compare the implementation to two algorithms available in SpaceEx~\cite{FrehseLGDCRLRGDM2011}, which is an efficient and mature verification tool for linear hybrid systems.
We summarize the different approaches below.

\textit{Deco.} This algorithm implements the decomposed approach presented in this paper. To compute the (low-dimensional) intersections, we use a polyhedra library~\cite{Polyhedra}.

\textit{LazySupp.} This algorithm uses a (lazy) support-function-based approximation of the intersection operation using the simple heuristic $\rho_{X \cap Y}(\ell) \leq \min({\rho_X(\ell), \rho_Y(\ell)})$.
This heuristic is fast but not precise enough to verify the safety property of the \emph{filtered oscillator} model.

\textit{LazyOptim.} In contrast to the coarse intersection in LazySupp, this algorithm uses a more precise implementation of the intersection operation based on line search~\cite{FrehseR12}.

\textit{SpaceEx LGG.} This is an efficient implementation of the algorithm by Le Guernic and Girard~\cite{LeGuernicG09}.
The algorithm uses a support-function representation and can hence check low-dimensional conditions efficiently, but operations such as intersections work in high dimensions.

\textit{SpaceEx STC.} The STC algorithm is an extension of the LGG algorithm with automatic time-step adaptation~\cite{FrehseKG13}.

\medskip

All tools use a support-function representation of sets.
We use template polyhedra with box constraints to overapproximate these sets, which roughly corresponds to a decomposition into one-dimensional blocks.
This approximation is fast.

For the algorithms implemented in JuliaReach, we use one-dimensional block structures for all models.
For SpaceEx we use the options given in the benchmark sources.
For algorithms that use a fixed time step, we fix this parameter to the same value for each benchmark.
The SpaceEx STC algorithm does not have such a parameter, so we instead fix the parameter ``flowpipe tolerance,'' which controls the approximation error, to the following values: \emph{Linear switching system}: $0.01$, \emph{Spacecraft rendezvous}: $0.2$, \emph{Platooning}: $1$, \emph{Filtered oscillator}: $0.05$.

\subsection{Experimental results}

The results are presented in Table~\ref{tab:benchmarks}.
We generally observe a performance boost of the Deco algorithm for models with more than five dimensions, and only a minor overhead for ``small'' models.
This demonstrates the general scalability improvement by performing operations, especially the intersection, in low dimensions.
Since all models have a small number of constrained dimensions in their guards and invariants, choosing an appropriate block structure results in very low-dimensional sets for computing the intersections, for which concrete polyhedral computations are very efficient and most precise.
We note that such an intersection computation does not scale with the dimension, and so other algorithms must resort to approximation techniques.

\begin{table}[t]
	\caption{
	Evaluation for different time steps on the \emph{filtered oscillator} model with 64 filters (``filtered\_osc64'' in Table~\ref{tab:benchmarks}) and all algorithms (see Section~\ref{sec:tool_descriptions}) that allow varying the time step.
	The last row shows the relative change of the fourth row over the first row as base line.
	}
	\label{tab:fo_time_step}
	\centering
	\renewcommand{\arraystretch}{1.1}
	\def\tabspaceborder{\hspace*{1mm}}
	\def\tabspace{\hspace*{3mm}}
	\def\mcc#1{\multicolumn{1}{c @{\tabspace}}{#1}}
	\centering
	\begin{tabular}{@{\tabspaceborder} l @{\tabspace} c @{\tabspace} c @{\tabspace} c @{\tabspace} c @{\tabspaceborder}}
		\toprule
		Step        & Deco              & LazySupp          & LazyOptim         & SpaceEx LGG \\
		\midrule
		0.01        & $2.8 \times 10^0$ & $7.4 \times 10^0$ & $5.6 \times 10^2$ & $2.0 \times 10^1$ \\
		0.005       & $4.1 \times 10^0$ & $1.3 \times 10^1$ & $9.5 \times 10^2$ & $3.6 \times 10^1$ \\
		0.001       & $1.7 \times 10^1$ & $6.5 \times 10^1$ & $4.7 \times 10^3$ & $1.6 \times 10^2$ \\
		0.0005      & $3.2 \times 10^1$ & $1.3 \times 10^2$ & $8.4 \times 10^3$ & $3.2 \times 10^2$ \\
		\midrule
		$\times 20.0$ & $\times 11.4$   & $\times 17.5$    & $\times 15.0$    & $\times 16.0$ \\
		\bottomrule
	\end{tabular}
\end{table}

Moreover, we found that our approach scales more favorably compared to the high-dimensional approaches when decreasing the time step $\delta$ (cf.\ Section~\ref{sec:reach_lti}).
We demonstrate this observation for the \emph{filtered oscillator} model in Table~\ref{tab:fo_time_step} and explain it as follows.
Recall that we only compute those sets in the flowpipe in high dimensions for which we have detected an intersection in low dimensions.
With a smaller time step, the total number of sets increases and hence the savings due to our approach become more dominant.
To give an example, for the \emph{filtered oscillator} with time step $0.0005$, out of the $9{,}661$ sets in total we only computed $1{,}400$ sets in high dimensions.
Consequently, making the time step $20$ times smaller only makes the run time $11.4$ times slower for our algorithm as opposed to at least $15$ times slower for other algorithms.

\begin{figure}[tb]
	\centering
	\includegraphics[height=65mm,width=\textwidth,keepaspectratio]{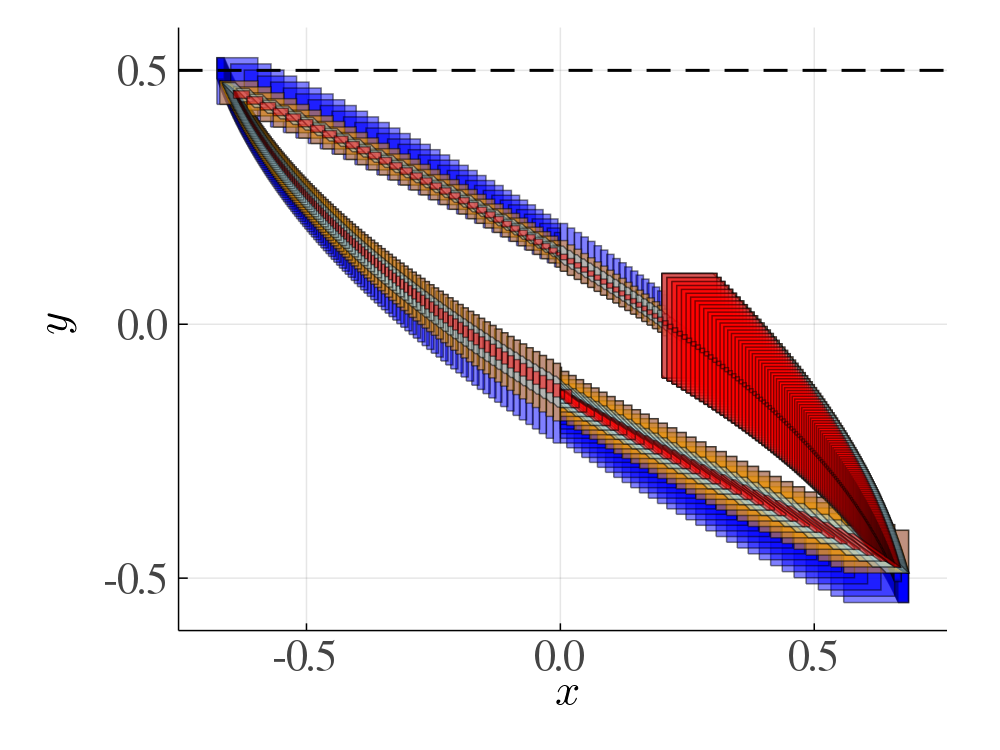}
	\caption{Four flowpipes for the model \emph{filtered oscillator} using a time step of $0.01$.
	The dark blue flowpipe (low-dimensional intersection algorithm) and the orange flowpipe (medium-dimensional intersection algorithm) are obtained for a one-dimensional block structure.
	The light blue flowpipe is obtained for a two-dimensional block structure and octagon approximation.
	The red flowpipe is obtained by the tool SpaceEx (which uses high-dimensional analysis).
	The dashed line corresponds to the safety property $y < 0.5$.}
	\label{fig:filtered_oscillator}
\end{figure}

\medskip

Furthermore, while in general our algorithm may induce an additional approximation error with the block structure, in the evaluation it was always precise enough to prove the safety properties. In Table~\ref{tab:benchmarks} we report the total amount of constrained dimensions in guards, invariants, and safety properties for each benchmark instance. These are the dimensions that determine our low-dimensional flowpipes. For some of the models, the constrained dimensions differ between the invariants/guards and the safety property, but a one-dimensional block structure was still sufficient.
Note that we need to compute intersections only with invariants and guards; for safety properties we just need to check inclusion.

In the \emph{linear switching} model, only one dimension is constrained, so our algorithm, together with one-dimensional block structure, is appropriate, especially because a small time step is required to prove the property.
In the \emph{platooning} model, the safety property constrains three dimensions but the invariants and guards constrain just one dimension.

\smallskip

The invariants and guards of the models \emph{spacecraft} and \emph{filtered oscillator} constrain two dimensions, so the natural choice for the decomposition is to keep these dimensions in the same block.
However, we chose to decompose into one-dimensional blocks for better run-time comparison in Table~\ref{tab:benchmarks}.
In the implementation, we follow the strategy to perform the intersection in two dimensions and then project back to one dimension (cf.\ Section~\ref{sec:deco_intersection}).
If instead we decide to employ a two-dimensional block structure, we can further gain precision by using different template polyhedra, e.g., with octagon constraints.
In Figure~\ref{fig:filtered_oscillator} we visualize this alternative for the \emph{filtered oscillator}.
As expected, the one-dimensional analysis is less precise as it corresponds to a box approximation.
However, we note that a one-dimensional block structure also inherently reduces the precision of \dwelldeco, so the additional approximation error does not only stem from the handling of discrete transitions by to our approach.
In addition, one can observe that the approximation using a one-dimensional block structure with low-dimensional intersection is coarser compared with the medium-dimensional strategy and we cannot prove the safety property anymore.

\subsection{Scaling the number of constrained dimensions}

We now investigate the effect of increasing the number of constrained dimensions in two experiments.

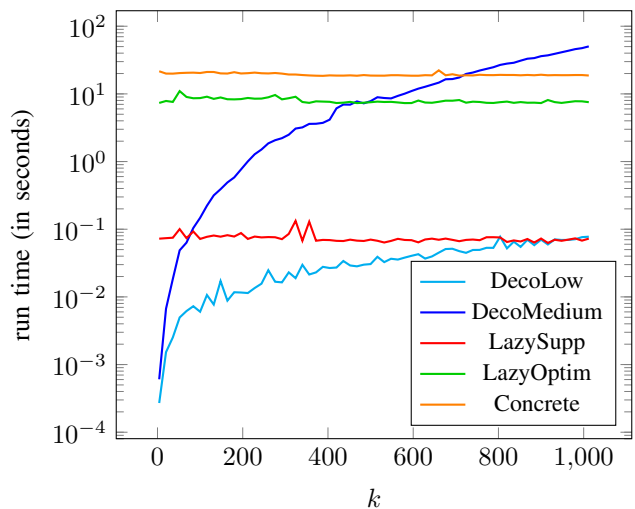
\begin{figure}[tb]
	\centering
	\input{experiment_intersection}
	\caption{Scaling the number of constrained dimensions $k$ for an intersection $\X \cap \YG$ where $\X$ is a hypercube and $\YG$ is a half-space.}
	\label{fig:scaling_intersection}
\end{figure}

In the first experiment we focus on the intersection operation.
We fix a hypercube $\X$ of dimension $1{,}024$, of radius~$4$ and centered in the origin, and a half-space $\YG$ $x_1 + \dots + x_k \leq 2$ for parameter~$k$ that controls the number of constrained dimensions.
The half-space properly cuts the hypercube for any~$k$.
In Figure~\ref{fig:scaling_intersection} we show the run times for different intersection algorithms.
We note that these algorithms compute different approximations of the true intersection:
the ``DecoLow'' and ``DecoMedium'' algorithms implement the respective ideas from Figure~\ref{fig:interectoin_algorithms}; the ``LazySupp'' and ``LazyOptim'' algorithms use the ideas as in the reachability algorithms of the same name (described in Section~\ref{sec:tool_descriptions}); the ``Concrete'' algorithm uses a polyhedra library (note that the H-representation of a hypercube of dimension $n$ has only $2n$ constraints).
We can see that the high-dimensional intersection algorithms are not affected by the number of constrained dimensions $k$ whereas the decomposition-based algorithms scale gracefully with $k$.

\begin{figure}[tb]
	\centering
	\input{experiment_reach_scale}
	\caption{Scaling the number of constrained dimensions $k$ for the modified \emph{filtered\_osc128} benchmark.}
	\label{fig:scaling_reach}
\end{figure}
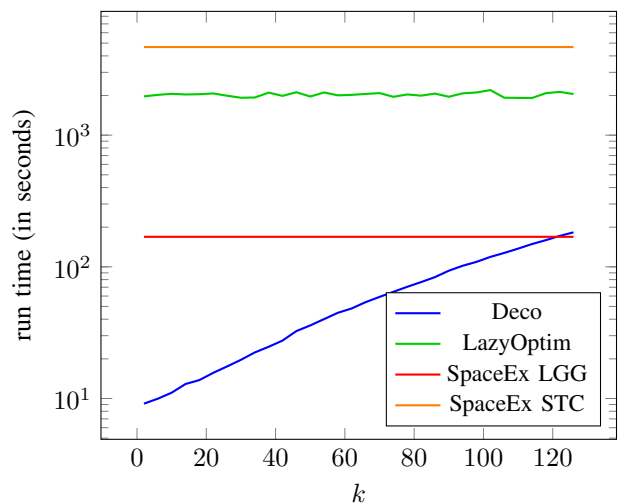

In the second experiment we consider the full reachability setting.
For that we modify the \emph{filtered oscillator} benchmark with $128$ filters (\emph{filtered\_osc128} in Table~\ref{tab:benchmarks}) by adding small nonzero entries to $k$ previously unconstrained dimensions in the invariants and guards.
We consider all the reachability algorithms that are precise enough to verify the safety property (which is still satisfied in our modified benchmark instances), i.e., all the algorithms from Table~\ref{tab:benchmarks} except for \emph{LazySupp}. Note that for SpaceEx algorithms we extrapolate the results we report for this benchmark instance in Table~\ref{tab:benchmarks} to a range of values of parameter $k$ as the reachability algorithms implemented in SpaceEx do not depend on the number of constrained dimensions. The run times are plotted in Figure~\ref{fig:scaling_reach} as a function of~$k$.
Again the high-dimensional algorithms are not affected by the number $k$ whereas the decomposition-based algorithm scales well with $k$ (note the log scale).
For $k \approx 120$ (close to $n = 131$) we see a cross-over with the \emph{SpaceEx LGG} run time, which is expected due to the overhead of the decomposition.

%% file: experiment_intersection.tex
\begin{tikzpicture}[scale=1]
	\begin{semilogyaxis}[legend style={legend pos=south east,font=\small},xlabel=$k$,ylabel=run time (in seconds)]
		\addplot[cyan,thick] table [x=k,y=DecoLow,col sep=comma,mark=none] {experiment_intersection_data.csv};
		\addlegendentry{DecoLow}
		\addplot[blue,thick] table [x=k,y=DecoMedium,col sep=comma,mark=none] {experiment_intersection_data.csv};
		\addlegendentry{DecoMedium}
		\addplot[red,thick] table [x=k,y=LazySupp,col sep=comma,mark=none] {experiment_intersection_data.csv};
		\addlegendentry{LazySupp}
		\addplot[green!80!black,thick] table [x=k,y=LazyOptim,col sep=comma,mark=none] {experiment_intersection_data.csv};
		\addlegendentry{LazyOptim}
		\addplot[orange,thick] table [x=k,y=Concrete,col sep=comma,mark=none] {experiment_intersection_data.csv};
		\addlegendentry{Concrete}
	\end{semilogyaxis}
\end{tikzpicture}

%% file: experiment_reach_scale.tex
\begin{tikzpicture}[scale=1]
	\begin{semilogyaxis}[legend style={legend pos=south east,font=\small},xlabel=$k$,ylabel=run time (in seconds)]
		\addplot[blue,thick] table [x=k,y=Deco,col sep=comma,mark=none] {experiment_reach_scale_data.csv};
		\addlegendentry{Deco}
		\addplot[green!80!black,thick] table [x=k,y=LazyOptim,col sep=comma,mark=none] {experiment_reach_scale_data.csv};
		\addlegendentry{LazyOptim}
		\addplot[red,thick] table [x=k,y=SpaceEx LGG,col sep=comma,mark=none] {experiment_reach_scale_data.csv};
		\addlegendentry{SpaceEx LGG}
		\addplot[orange,thick] table [x=k,y=SpaceEx STC,col sep=comma,mark=none] {experiment_reach_scale_data.csv};
		\addlegendentry{SpaceEx STC}
	\end{semilogyaxis}
\end{tikzpicture}

%% file: conclusion.tex
\section{Conclusion}\label{sec:conclusion}

We have presented a schema that integrates a reachability algorithm based on decomposition for LTI systems in the analysis loop for linear hybrid systems.
The key insight is that intersections with polyhedral constraints can be efficiently detected and computed (approximately or often even exactly) in low dimensions.
This enables the systematic focus on appropriate subspaces and the potential for bypassing large amounts of flowpipe computations.
Moreover, working with sets in low dimensions allows to use precise polyhedral computations that are infeasible in high dimensions.

\paragraph*{Future work}

An essential step in our algorithm is the fast computation of a low-dimensional flowpipe for the detection of intersections.
In the presented algorithm, we recompute the flowpipe for the relevant time frames in high dimensions using the same decomposed algorithm with the same time step.
However, this is not necessary.
We could achieve higher precision by using different algorithmic parameters or even a different, possibly non-decomposed, algorithm (e.g., one that features arbitrary precision~\cite{GirardLG08lgg}).
This is particularly promising for LTI systems because one can avoid recomputing the homogeneous (state-based) part of the flowpipe~\cite{FrehseR12}.

In the benchmarks considered in the experimental evaluation, it was not necessary to change the block structure when switching between locations.
In general, different locations may constrain different dimensions, so tracking the ``right'' dimensions may be necessary to maintain precision.
While it is easy to merge different blocks, subsequent computations would become more expensive.
Hence one may also want to split blocks again for optimal performance.
Since this comes with a loss in precision, heuristics for rearranging the block structure, possibly in a refinement loop, are needed.

%% file: proofs.tex
\appendix\label{sec:proofs}

\begin{IEEEproof}[Proof of Proposition~\ref{prop:error_decomposition}]
	Recall that
	\begin{align*}
		& \hausdorff(\X, \bigtimesdm_j \pi_j \X)
		% Hausdorff distance of nonempty compact convex sets
		\\
		={}& \inf_{\varepsilon \in \R} \left\{ \bigtimesdm_j \pi_j \X \subseteq \X \oplus \varepsilon\Bpn \text{ and } \X \subseteq \bigtimesdm_j \pi_j \X \oplus \varepsilon\Bpn \right\}.
	\end{align*}
	We prove that \pnorm{r_{\X}^p} is an upper bound on $\varepsilon$ by showing the two inclusions.
	One inclusion is trivial because $\X \subseteq \bigtimes_j \pi_j \X$.
	For the other inclusion, observe that the center of the box approximation of \X, $c_{\X}^p$, lies in \X.
	Recall that $p = \infty$ and that the box approximation of \X, $\{c_{\X}^p\} \oplus \pnorm{r_{\X}^p}\Bpn$, is the worst-case decomposition of \X, subsuming all other decompositions.
	\begin{equation*}
		\bigtimesdm_j \pi_j \X \subseteq \{c_{\X}^p\} \oplus \pnorm{r_{\X}^p}\Bpn \subseteq \X \oplus \pnorm{r_{\X}^p}\Bpn
		\tag*{\IEEEQEDhere}
	\end{equation*}
\end{IEEEproof}

\bigskip

\begin{IEEEproof}[Proof of Proposition~\ref{prop:intersection}]
	\begin{align*}
		& \hausdorff(\Xhat \cap \YG, \Xhat \cap \YGhat) \\
		% Hausdorff distance of nonempty compact convex sets
		={}& \max_{\pnorm{d} \leq 1} \rho_{\Xhat \cap \YGhat}(d) - \rho_{\Xhat \cap \YG}(d) \\
		% definition of support function of compact convex sets
		={}& \max_{\pnorm{d} \leq 1} \max_{x \in \Xhat \cap \YGhat} \dotp{d}{x} - \max_{y \in \Xhat \cap \YG} \dotp{d}{y} \\
		% pull out max
		={}& \max_{\pnorm{d} \leq 1} \max_{x \in \Xhat \cap \YGhat} \min_{y \in \Xhat \cap \YG} \dotp{d}{x - y} \\
		\intertext{Choosing $d$ accordingly we take absolutes in the dot product.}
		={}& \max_{\pnorm{d} \leq 1} \max_{x \in \Xhat \cap \YGhat} \min_{y \in \Xhat \cap \YG} \dotp{d}{|x - y|} \\
		% enlarge domain for y
		\leq{}& \max_{\pnorm{d} \leq 1} \max_{x, y \in \Xhat \cap \YGhat} \dotp{d}{|x - y|} \\
		\intertext{We conservatively bound $|x - y|$ by the diameter of $\Xhat \cap \YGhat$.}
		\leq{}& \max_{\pnorm{d} \leq 1} \dotp{d}{\Delta_p(\Xhat \cap \YGhat)} \\
		%
		% estimate diameter by individual diameters
		\leq{}& \min\left( \max_{\pnorm{d} \leq 1} \dotp{d}{\Delta_p(\Xhat)}, \max_{\pnorm{d} \leq 1} \dotp{d}{\Delta_p(\YGhat)} \right) \\
		% dot product
		\leq{}& \max_j \min(\pnorm{\Delta_p(\X_j)}, \pnorm{\Delta_p(\pi_j \YG)})
		\tag*{\IEEEQEDhere}
	\end{align*}
\end{IEEEproof}

\begin{figure}[tb]
	\centering
	\includegraphics[height=35mm,width=\textwidth,keepaspectratio]{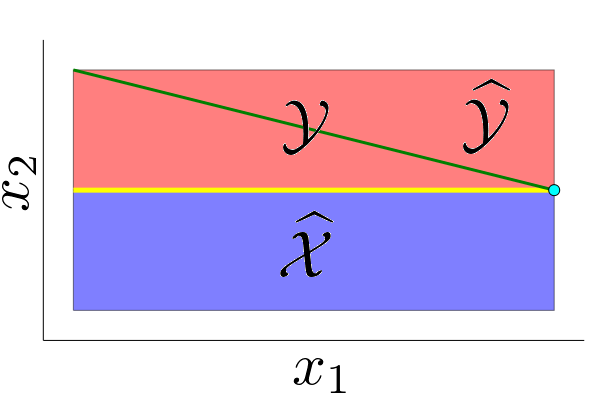}
	\caption{The rectangle \Xhat (blue) and the line segment \YG (green) intersect in a single point (cyan).
	The Cartesian decomposition \YGhat of \YG (red) intersects with \Xhat on a whole facet (yellow).
	Observe that the length of this intersection facet is determined by the width of \Xhat and \YG in dimension $x_1$, independent of the height in dimension $x_2$.}
	\label{fig:tight_intersection}
\end{figure}

Figure~\ref{fig:tight_intersection} shows that the bound of Proposition~\ref{prop:intersection} is tight.

\bigskip

\begin{IEEEproof}[Proof of Proposition~\ref{prop:inclusion}]
	\begin{align*}
		& \Xhat \subseteq \YGhat
		\iff \forall x \in \Xhat: x \in \YGhat \\
		\iff& \forall x_1 \in \X_1, \dots, x_\bl \in \X_\bl: \bigwedge_{j=1}^\bl x_j \in \YG_j \\
		\iff& \bigwedge_{j=1}^\bl \forall x_j \in \X_j: x_j \in \YG_j
		\iff \bigwedge_{j=1}^\bl \X_j \subseteq \YG_j
		\tag*{\IEEEQEDhere}
	\end{align*}
\end{IEEEproof}

\bigskip

\begin{IEEEproof}[Proof of Proposition~\ref{prop:convex_hull}]
	Let $d \in \R^n$.
	We show $\rho_{\CH(\Xhat \cup \YGhat)}(d) \leq \rho_{\bigtimes_j \CH(\X_j \cup \YG_j)}(d)$.
	\begin{align*}
		&\rho_{\CH(\Xhat \cup \YGhat)}(d)
		= \max(\rho_{\Xhat}(d), \rho_{\YGhat}(d)) \\
		={}& \max\left(\sumdm_j \rho_{\X_j}(d_j), \sumdm_j \rho_{\YG_j}(d_j)\right) \\
		\leq{}& \sumdm_j \max(\rho_{\X_j}(d_j), \rho_{\YG_j}(d_j)) \\
		={}& \sumdm_j \rho_{\CH(\X_j \cup \YG_j)}(d_j)
		= \rho_{\bigtimesdm_j \CH(\X_j \cup \YG_j)}(d)
		\tag*{\IEEEQEDhere}
	\end{align*}
\end{IEEEproof}

\bigskip

\begin{IEEEproof}[Proof of Proposition~\ref{prop:convex_hull_error}]
	The first bound is due to Proposition~\ref{prop:error_decomposition}.
	\begin{align*}
		& \hausdorff(\CH(\Xhat \cup \YGhat), \bigtimesdm_j \CH(\X_j \cup \YG_j)) \\
		% Hausdorff distance of nonempty compact convex sets
		={}& \max_{\pnorm{d} \leq 1} \rho_{\bigtimesdm_j \CH(\X_j \cup \YG_j)}(d) - \rho_{\CH(\Xhat \cup \YGhat)}(d) \\
		% simplification rules for support function
		={}& \max_{\pnorm{d} \leq 1} \left( \sumdm_j \rho_{\CH(\X_j \cup \YG_j)}(d_j) \right) - \max(\rho_{\Xhat}(d), \rho_{\YGhat}(d)) \\
		% simplification rules for support function; unroll structure of \Xhat/\YGhat
		={}& \max_{\pnorm{d} \leq 1} \left( \sumdm_j \underbrace{\max(\rho_{\X_j}(d_j), \rho_{\YG_j}(d_j))}_{=: \varphi(d_j)} \right) - {} \\
		& \quad \max\left( \sumdm_j \rho_{\X_j}(d_j), \sumdm_j \rho_{\YG_j}(d_j) \right) \\
		% pull out max (invert to min); combine block terms
		={}& \max_{\pnorm{d} \leq 1} \min\left( \sumdm_j \varphi(d_j) - \rho_{\X_j}(d_j), \sumdm_j \varphi(d_j) - \rho_{\YG_j}(d_j) \right) \\
		% re-substitute phi and simplify
		={}& \max_{\pnorm{d} \leq 1} \min\left(
			\begin{tabular}{l}
				$\sumdm_j \max(0, \rho_{\YG_j}(d_j) - \rho_{\X_j}(d_j))$, \\
				$\sumdm_j \max(0, \rho_{\X_j}(d_j) - \rho_{\YG_j}(d_j))$
			\end{tabular}
			\right) \\
		% replace min by absolute difference
		\leq{}& \max_{\pnorm{d} \leq 1} \sumdm_j |\rho_{\X_j}(d_j) - \rho_{\YG_j}(d_j)|
		\tag*{\IEEEQEDhere}
	\end{align*}
\end{IEEEproof}